\documentclass[iop,twocolappendix]{emulateapj}

\usepackage{color}
\usepackage{natbib}
\usepackage{graphicx}
\usepackage{amsmath}
\usepackage{collectbox}
\newcommand{\mini}{\mbox{$M_{\rm i}$}}
\newcommand{\Mini}{\mbox{$M_{\rm i}$}}
\newcommand{\Mcore}{\mbox{$M_{\rm core}$}}

\newcommand{\sub}[1]{\mbox{$_{\rm #1}$}}

\newcommand{\jh}{\mbox{$J\!-\!H$}}

\newcommand{\jks}{\mbox{$J\!-\!K_{\rm s}$}}

\newcommand{\ks}{\mbox{$K_{\rm s}$}}

\newcommand{\mh}{\mbox{\rm [{\rm M}/{\rm H}]}}
\newcommand{\Msun}{\mbox{$M_{\odot}$}}

\newcommand{\Mi}{\mbox{$M\sub{i}$}}
\newcommand{\Zi}{\mbox{$Z\sub{i}$}}
\newcommand{\Teff}{\mbox{$T_{\rm eff}$}}

\newcommand{\logL}{\mbox{$\log L/L_{\odot}$}}
\newcommand{\logte}{\mbox{$\log\Teff$}}

\newcommand{\logg}{\mbox{$\log g$}}
\newcommand{\co}{\mbox{${\rm C/O}$}}
\newcommand{\nintp}{\mbox{$n_{\rm inTPC}$}}

\newcommand{\beq}{\begin{equation}}
\newcommand{\eeq}{\end{equation}}
\newcommand{\beqa}{\begin{eqnarray}}
\newcommand{\eeqa}{\end{eqnarray}}

\newcommand{\citeinprep}[1]{#1 (in prep.)}

\newcommand{\parsec}{\texttt{PARSEC}}
\newcommand{\param}{\texttt{PARAM}}
\newcommand{\colibri}{\texttt{COLIBRI}}
\newcommand{\trilegal}{\texttt{TRILEGAL}}
\newcommand{\aesopus}{\texttt{{\AE}SOPUS}}

\slugcomment{To appear soon in ApJ}

\shorttitle{\parsec--\colibri\ isochrones}
\shortauthors{Marigo et al.}

\begin{document}


\title{A new generation of \parsec--\colibri\ stellar isochrones including the TP-AGB phase}

\author{
Paola Marigo\altaffilmark{1}, 
L\'eo Girardi\altaffilmark{2}, 
Alessandro Bressan\altaffilmark{3}, 
Philip Rosenfield\altaffilmark{4}, 
Bernhard Aringer\altaffilmark{1},
Yang Chen\altaffilmark{1},
Marco Dussin\altaffilmark{1},
Ambra Nanni\altaffilmark{1},
Giada Pastorelli\altaffilmark{1},
Tha\'{\i}se S.\ Rodrigues\altaffilmark{1,2}, 
Michele Trabucchi\altaffilmark{1},
Sara Bladh\altaffilmark{1},
Julianne Dalcanton\altaffilmark{5},
Martin A.T.\ Groenewegen\altaffilmark{6},
Josefina Montalb\'an\altaffilmark{1},
Peter R.\ Wood\altaffilmark{7}
\altaffiltext{1}{Dipartimento di Fisica e Astronomia Galileo Galilei, Universit\`a di Padova, Vicolo dell'Osservatorio 3, I-35122 Padova, Italy}
\altaffiltext{2}{Osservatorio Astronomico di Padova -- INAF, Vicolo dell'Osservatorio 5, I-35122 Padova, Italy}
\altaffiltext{3}{SISSA, via Bonomea 365, I-34136 Trieste, Italy}
\altaffiltext{4}{Harvard-Smithsonian Center for Astrophysics, 60 Garden St., Cambridge, MA 02138, USA}
\altaffiltext{5}{Department of Astronomy, University of Washington, Box 351580, Seattle, WA 98195, USA}
\altaffiltext{6}{Koninklijke Sterrenwacht van Belgi\"e, Ringlaan 3, 1180 Brussel, Belgium}
\altaffiltext{7}{Research School of Astronomy and Astrophysics, Australian National University, Cotter Road, Weston Creek, ACT 2611, Australia}
}

\begin{abstract}
We introduce a new generation of \parsec--\colibri\ stellar isochrones that include a detailed treatment of the thermally-pulsing asymptotic giant branch (TP-AGB) phase, and covering a wide range of initial metallicities ($0.0001 < \Zi < 0.06$). Compared to previous releases, the main novelties and improvements are: use of new TP-AGB tracks and related atmosphere models and spectra for M and C-type stars; inclusion of the surface H+He+CNO abundances in the isochrone tables, accounting for the effects of diffusion, dredge-up episodes and hot-bottom burning; inclusion of complete thermal pulse cycles, with a complete description of the in-cycle changes in the stellar parameters; new pulsation models to describe the long-period variability in the fundamental and first overtone modes; new dust models that follow the growth of the grains during the AGB evolution, in combination with radiative transfer calculations for the reprocessing of the photospheric emission. Overall, these improvements are expected to lead to a more consistent and detailed description of properties of TP-AGB stars expected in resolved stellar populations, especially in regard to their mean photometric properties from optical to mid-infrared wavelengths. We illustrate the expected numbers of TP-AGB stars of different types in stellar populations covering a wide range of ages and initial metallicities, providing further details on the ``C-star island'' that appears at intermediate values of age and metallicity, and about the AGB-boosting effect that occurs at ages close to 1.6-Gyr for populations of all metallicities. The isochrones are available through a new dedicated web server. 

\end{abstract}




\section{Introduction}
\label{intro}

Theoretical stellar isochrones are remarkably useful datasets in astrophysics. Historically, they have been commonly used to attribute approximated ages and distances to star clusters observed in at least two filters, in the traditional process of isochrone fitting \citep[since e.g.][]{demarque77}. In the last decades, their use has expanded in many different ways. Since \citet{charlot91}, they are at the basis of the isochrone method of the spectrophotometric evolutionary population synthesis, which has revolutionized the interpretation of the spectra and photometry of distant galaxies. They have also been extensively used in the more complex methods of CMD fitting of nearby galaxies and star clusters \citep[e.g.][]{dolphin02}, which aim at deriving quantitative measurements of their star formation and chemical enrichment histories. 

While the most traditional applications of isochrones refer to the interpretation of spectrophotometric data only, more recent applications regard other quantities as well. For instance, the advent of large microlensing \citep{paczynski96} and spectroscopic surveys \citep{york00} in the nineties has opened access to the variability and the chemical composition information for huge populations of stars in the Milky Way and its satellite galaxies, while the onset of space-based asteroseismology now gives access to fundamental quantities such as the masses and radii even for stars located tens of kiloparsecs away \citep{chaplin13}. The interpretation of such data in terms of stellar populations and ages requires that, besides the photometry, additional intrinsic stellar properties are provided along the isochrones.  

This need for more information is particularly important when we refer to thermally-pulsing asymptotic giant branch (TP-AGB) stars. They contribute to a sizable fraction of the integrated light of stellar populations, and are responsible for a significant fraction of the chemical enrichment and dust production in galaxies. Yet, the TP-AGB phase is still the most uncertain among the main evolutionary phases of single stars, since its evolution is determined by a series of complex and interconnected processes that still cannot be simply modeled from first principles -- like the third dredge-up episodes, long-period pulsation and mass loss, etc.\ \citep[see][]{frost96, mowlavi99, herwig05, marigo15}. The calibration of this phase {\it requires} the construction of extended sets of isochrones in which all the relevant processes are considered and all the relevant observables are tabulated. Only in this way we can enable a quantitative comparison between the model predictions and the properties of TP-AGB populations observed in nearby galaxies.

Indeed, in \citet{marigo08} we made a significant step in this direction, by providing the first isochrones in which the properties of TP-AGB stars were considered in more detail, and including crucial processes like third dredge-up, hot-bottom burning, low-temperature opacity changes, the distinct spectra of carbon (C) stars, the reprocessing of radiation by circumstellar dust in phases of increased mass-loss, and the expected pulsation periods. In this paper, we provide new isochrones derived from the stellar evolutionary tracks computed with the more recent \parsec\ \citep{bressan12} and \colibri\ \citep{marigo13} codes. They include more details than the previous isochrones, and are specifically designed to allow us to advance in the process of calibration of the TP-AGB evolution. Some aspects of these isochrones, such as the inclusion of chemical abundance information, will be of interest to many other isochrone users as well. The input data and methods are described in Sect.~\ref{datamethods}. Some new properties of the new isochrones are illustrated in Sect.~\ref{results}. Data retrieval and ongoing work are briefly described in Sect.~\ref{conclu}.

\section{Data and methods}
\label{datamethods}

\subsection{\parsec\ tracks}

\parsec\ (the PAdova and tRieste Stellar Evolutionary Code) represents a deeply-revised version of the Padova code used in many popular sets of isochrones \citep[e.g.\ in][]{bertelli94, girardi00, girardi10, marigo08}. The main features in \parsec\ are described in \citet{bressan12,bressan13}, and include the updating of the input physics (equation of state, opacities, nuclear reaction rates, solar reference abundances) and of mixing processes, in particular with the addition of microscopic diffusion in low-mass stars. Further updates regard the treatment of boundary conditions in low-mass stars \citep{chen14}, and envelope overshooting and mass-loss in intermediate- and high-mass stars \citep{tang14, chen15}. As it is clear from these papers, \parsec\ is a rapidly-evolving code, with further revisions being underway. 

\begin{figure*}
\includegraphics[width=\textwidth]{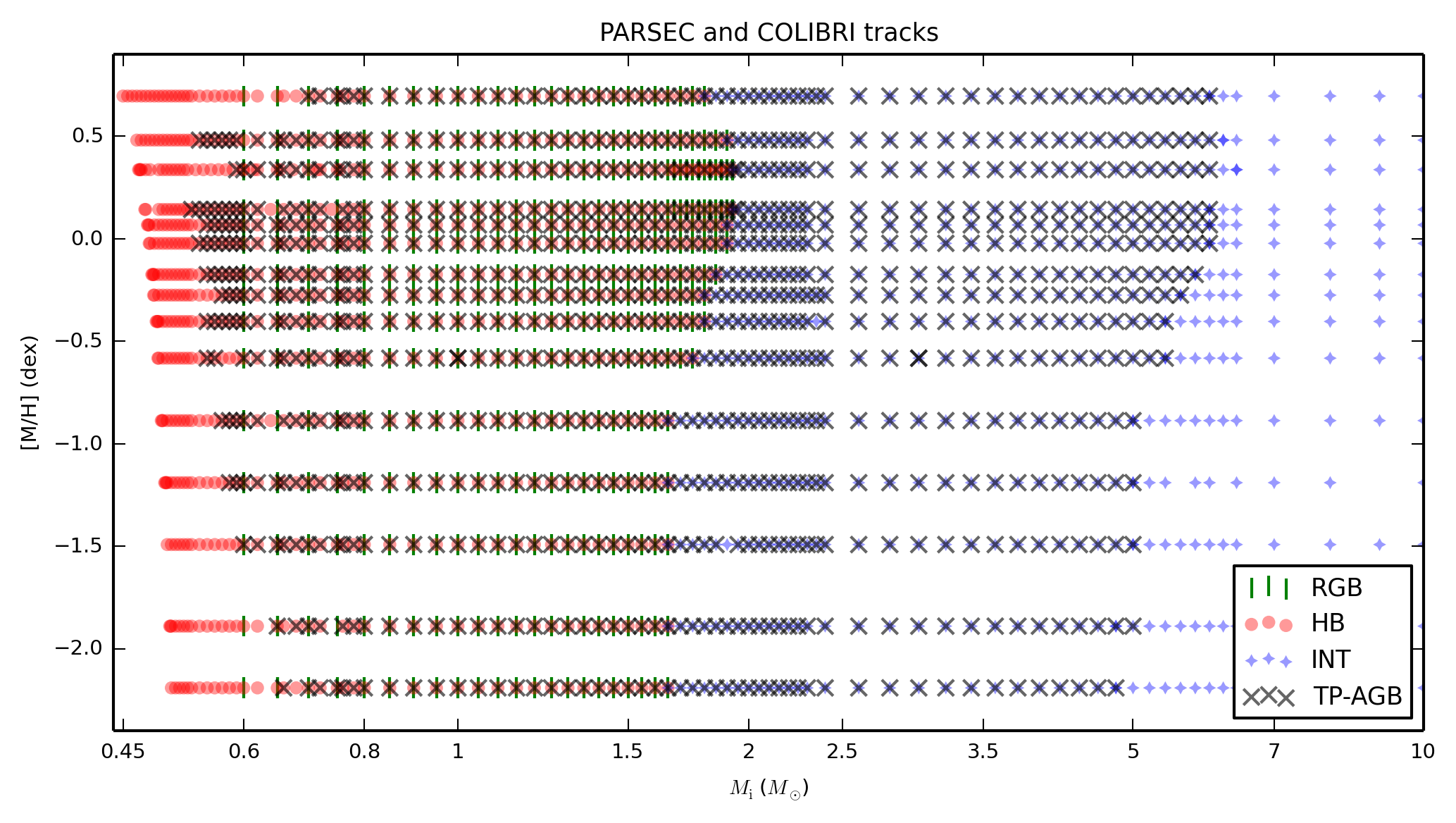}
\caption{The set of tracks involved in the construction of \parsec-\colibri\ isochrones,  in the \mh\ vs.\ \mini\ plane. Tracks from \parsec\ are divided into three broad classes tagged as RGB (low-mass tracks from the pre-MS up to the He-flash at the tip of the RGB), HB (low-mass tracks from the start of quiescent core He-burning up to the first TPC on the TP-AGB) and INT (intermediate-mass and massive tracks from the pre-MS up to either a first TPC, or C-ignition). For all cases where a first TPC was detected, the TP-AGB evolution (as tagged) was followed with the \colibri\ code. In addition to the tracks here presented, the \parsec\ database includes very low-mass tracks down to 0.1~\Msun\ evolved up to an age larger than the Hubble time, and massive tracks up to 350~\Msun\ evolved up to C ignition; they are not shown in the plot but are also included in the available isochrones.}
\label{fig:track_masses}
\end{figure*}

In the present work, we use a subset of the \parsec\ V1.2S evolutionary tracks as a reference. They include grids of tracks for 15 values of initial metal content, $Z_\mathrm{i}$, between 0.0001 and 0.06. The helium initial content, $Y_\mathrm{i}$, follows the initial metal content according to the $Y_\mathrm{i}=1.78\times Z_\mathrm{i}+0.2485$ relation, so as to reproduce both the primordial $Y$ value by \citet{komatsu11}, and the chemical composition of the present Sun \citep[namely $Z_\odot=0.01524$, $Y_\odot=0.2485$, see][for details]{bressan13}. The reference solar-scaled composition is taken from \citet{caffau11}. Adopting the simple approximation of $\mh=\log(Z_\mathrm{i}/Z_\odot)-\log(X_\mathrm{i}/X_\odot)$, the metallicity \mh\ ranges from $-2.19$ to $+0.70$~dex. 

The range of masses computed with \parsec\ is also very wide, generally including $\sim\!120$ different mass values distributed in the range from 0.1 to 350~\Msun, for each metallicity. The mass range more relevant for this paper is the one from $\sim0.5$~\Msun\ up to the maximum mass of stars developing the TP-AGB phase, which is located somewhere between 5 and 6.4~\Msun, depending on metallicity. This mass--metallicity range is presented in Fig.~\ref{fig:track_masses}, together with a summary of the evolutionary phases computed in every case.

\subsection{\colibri\ tracks}

The TP-AGB evolutionary tracks from the first thermal pulse up to the complete ejection of the envelope are computed with the \colibri\ code developed by \citet{marigo13}, and including the updates in the mass-loss described by \citet{rosenfield14, rosenfield16}. While many details of \colibri\ are described in the aforementioned works, let us highlight here the main differences with respect to the previous TP-AGB tracks computed by \citet{marigo07}, which were used in the stellar isochrones of \citet{marigo08}. We recall that \colibri\ couples a synthetic module (which contains the main free parameters to be calibrated) with the numerical solution of the atmosphere and complete envelope model, down to the bottom of the H-burning shell. The integration strategy is detailed \citet{marigo13}, where several accuracy tests with respects to full stellar models are also presented.  \colibri\ is characterised by a high computational speed and a robust numerical stability, allowing us to promptly compute large sets of TP-AGB tracks any time we aim at exploring the impact of a different model prescription.  Such a feature is really essential to perform the demanding TP-AGB calibration cycle. At the same time,  \colibri\ incorporates many revisions in the input physics, some of which (e.g.\ the nuclear reaction rates) are also common to \parsec. More importantly, \colibri\ is the first code to fully include the on-the-fly computation of the equation of state and Rosseland mean opacities 
by suitably calling, at each time step,  
the \texttt{Opacity Project} routines \citep[for $T> 12.000$K;][]{Seaton05}, and the \aesopus\ routines \citep[for $1.500 \le T \le 12.000$K;][]{marigo09} in full consistency with the actual composition of the stellar envelope. Therefore, composition-changing processes such as the third-dredge up (3DU) and the hot-bottom burning (HBB) at the base of the convective envelope, produce an immediate effect in the stellar structure and on stellar properties such as the effective temperature, \Teff, hence also changing the efficiency of other processes such as the mass-loss rate, $\dot{M}$, pulsation periods, etc. Similar effects of feedback between chemical composition and stellar structure were also included in \citet{marigo07} models, but in a much simpler way \citep[see][]{marigo02}. Moreover, \colibri\ improves in the description of other effects, such as the stellar sphericity, and the integration of extended nuclear networks for all burning processes -- including the HBB which requires a detailed consideration of both nuclear and convective timescales.

The occurrence and efficiency of the 3DU process in TP-AGB stars is notoriously uncertain and sensitive to numerical details \citep{frost96, mowlavi99}. As in \citet{marigo07} models, this process is parameterised also in \colibri. This causes the model to have a few efficiency parameters (for 3DU and mass-loss) that can be tuned so as to reproduce the observed properties of populations of AGB stars.

The \colibri\ models illustrated in this work are the same ones as described in \citet{rosenfield16}, which are shown to reproduce the AGB star numbers in a set of nearby dwarf galaxies. They include the complete range of masses and metallicities required to complement the \parsec\ V1.2S grid of tracks, as shown in Fig.~\ref{fig:track_masses}. About $60$ TP-AGB tracks are computed for each metallicity. The initial conditions of every \colibri\ track (core mass, luminosity, envelope composition, etc.) are taken taken from the corresponding \parsec\ track just before the first significant thermal pulse; therefore the subsequent section of the \parsec\ track is simply replaced by the \colibri\ one. As illustrated by \citet{marigo13}, this gives origin to an almost continuous track, with just a small jump in the \Teff\ (typically less that 50~K, see their figure 6) at the \parsec--\colibri\ junction.

\begin{figure*}
\includegraphics[width=0.75\textwidth]{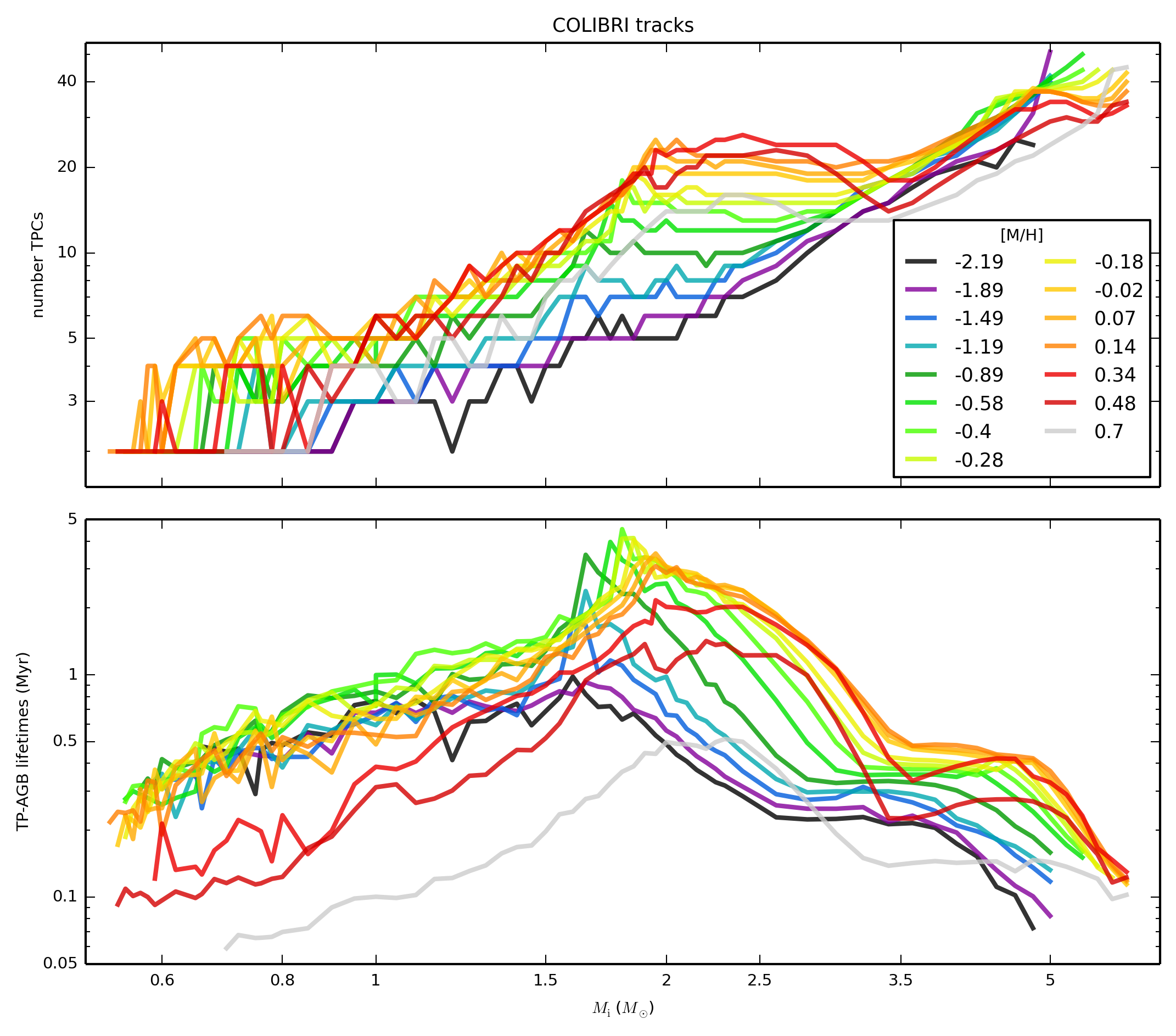}
\caption{Number of TPCs in every \colibri\ TP-AGB track (top panel), and its total lifetime (bottom panel).}
\label{fig:tracks_tpcs}
\end{figure*}

Figure~\ref{fig:tracks_tpcs} shows two of the basic properties of these tracks, namely the number of TPCs and the total TP-AGB lifetime, as a function of initial mass and metallicity.

\subsection{Equivalent evolutionary points and interpolation method}

The grids of stellar evolutionary tracks describe how the stellar properties -- here generically denoted by $p$ -- vary as a function of stellar age, $t$, for a given set of initial masses, \Mini. Building isochrones is essentially the process of interpolating inside this grid to produce a sequence of such properties as a function of \Mini, for a given set of $t$, that is, 
\begin{equation}
\boxed{\begin{array}{c}
 \mathrm{Tracks} \\
 p(t)|_{M_\mathrm{i}=\mathrm{const.}} 
\end{array} }
\Rightarrow 
\boxed{\begin{array}{c}
 \mathrm{Isochrones} \\
 p(M_\mathrm{i})|_{t=\mathrm{const.}}
\end{array} }
 \,\,\,\,. 
\end{equation}

For doing so, we use the classic method of interpolating between tracks using pairs of equivalent evolutionary points (EEP) as a reference. Essentially, once a pair of EEPs is identified in two adjacent tracks, the whole evolutionary sequence between these EEPs is assumed to be equivalent and interpolated in all quantities, by using \Mini\ and age as the independent variables. After building a dense grid of interpolated tracks of masses \Mini, the simple selection of points with desired age $t$ allows us to draw a complete isochrone. Moreover, by following an algorithm similar to \citet{bertelli08}, the available grids of tracks are also interpolated for any intermediate value of initial metallicity. Tracks that share similar evolutionary properties -- e.g. tracks with an extended red giant branch (RGB), or tracks with a convective core on the main sequence -- are grouped together and have their EEPs selected with exactly the same criteria, so that the interpolations turn out to be smooth not only as a function of initial mass and age, but also of metallicity. 

According to this scheme, the interpolated tracks can be built with any arbitrary density in initial mass coordinate; the higher the density, the larger the number of points in the resulting isochrones. In order to limit these numbers, we use a dynamical interpolation algorithm that, for every track section being considered, creates as many interpolated tracks (and hence isochrone points) as necessary to obtain a given resolution in the $\log L$ versus $\log\Teff$ space. In the present release, this resolution is set to a minimum of $\Delta\log L=0.04$~dex and $\Delta\log\Teff=0.01$~dex, for all sections of \parsec\ tracks; this results in isochrones containing a few hundred points before the first TPC. The TP-AGB resolution is defined in a different way, as we specify below.

The algorithm we developed to build the isochrones is very general, even allowing us to identify the presence of multiple sequences of evolutionary stages along the same isochrone, as those illustrated in \citet{girardi13}; more specifically, in that work we have identified the presence of double or triple TP-AGB sequences causing a marked ``AGB-boosting'' effect in isochrones (and star clusters) of ages close to 1.6-Gyr.  

\subsection{Mass loss on the RGB}

Another important detail is that we can apply a modest amount of mass loss between the tip of the RGB and the zero-age core-helium burning (CHeB) stage of low-mass stars, in order to simulate the effect of mass-loss along the RGB. Just as in previous releases \citep{girardi00,marigo08}, this is done in an approximative way: we first estimate the amount of mass loss expected from a given mass-loss formula, $\Delta M$, by integrating the mass-loss rate along the RGB section of the evolutionary tracks, i.e.\ $\Delta M=\int_\mathrm{RGB} \dot{M} \mathrm{d}t$. Then, the function $\Delta M(\Mini)$ is used to assign every RGB track of mass \Mini\ to a given CHeB+AGB evolutionary sequence of mass $\Mini-\Delta M(\Mini)$. The CHeB+AGB sequence is created by interpolation in the existing grid, and then attached to the RGB track while ensuring the continuity of the stellar age along the entire sequence. The modified RGB+CHeB+AGB tracks are then used to build the isochrones. This is a realistic approximation for low-mass stars, in which the RGB mass loss neither affect the evolution of the stellar core, nor the shape of evolutionary tracks in a significant way.

By default, we apply the classical \citet{reimers75}'s mass loss formula with a multiplicative coefficient of $\eta_\mathrm{R}=0.2$ \citep{miglio12},
for stars of masses smaller than 1~\Msun. Then, an additional multiplicative factor is assumed in order to decrease the mass loss predicted by this formula gradually as the initial mass increases from 1.0 to 1.7~\Msun. In this way, stars of masses $\Mi\ga1.5$~\Msun, for which the present algorithm might provide inconsistent results, are little affected by mass loss. Stars of masses $\Mi\ga1.7$~\Msun\ are assumed not to lose mass before the TP-AGB.
 
\subsection{Thermal pulse cycle $L$ variations}

Thermal pulse cycle (TPC) $L$ and \Teff\ variations are a basic feature of TP-AGB evolutionary tracks. They are quasi-periodic changes in $L$ and \Teff\ caused by the onset of He-shell flashes and the subsequent cycle of envelope expansion, switching off of the He-shell, contraction, and gradual expansion as the quiescent H-shell burning takes over \citep[see e.g.][]{boothroyd88, vassiliadis93, wagenhuber98}. These basic features of the tracks, however, are very hard to obtain in the isochrones. The main problem is that the number of TPCs in general varies from track to track, even if they are separated by small intervals of initial mass and metallicity, e.g. 0.05~\Msun\ and 0.1~dex, respectively. This variation can be appreciated in the top panel of Fig.~\ref{fig:tracks_tpcs}. 

\begin{figure*}
\includegraphics[width=\textwidth]{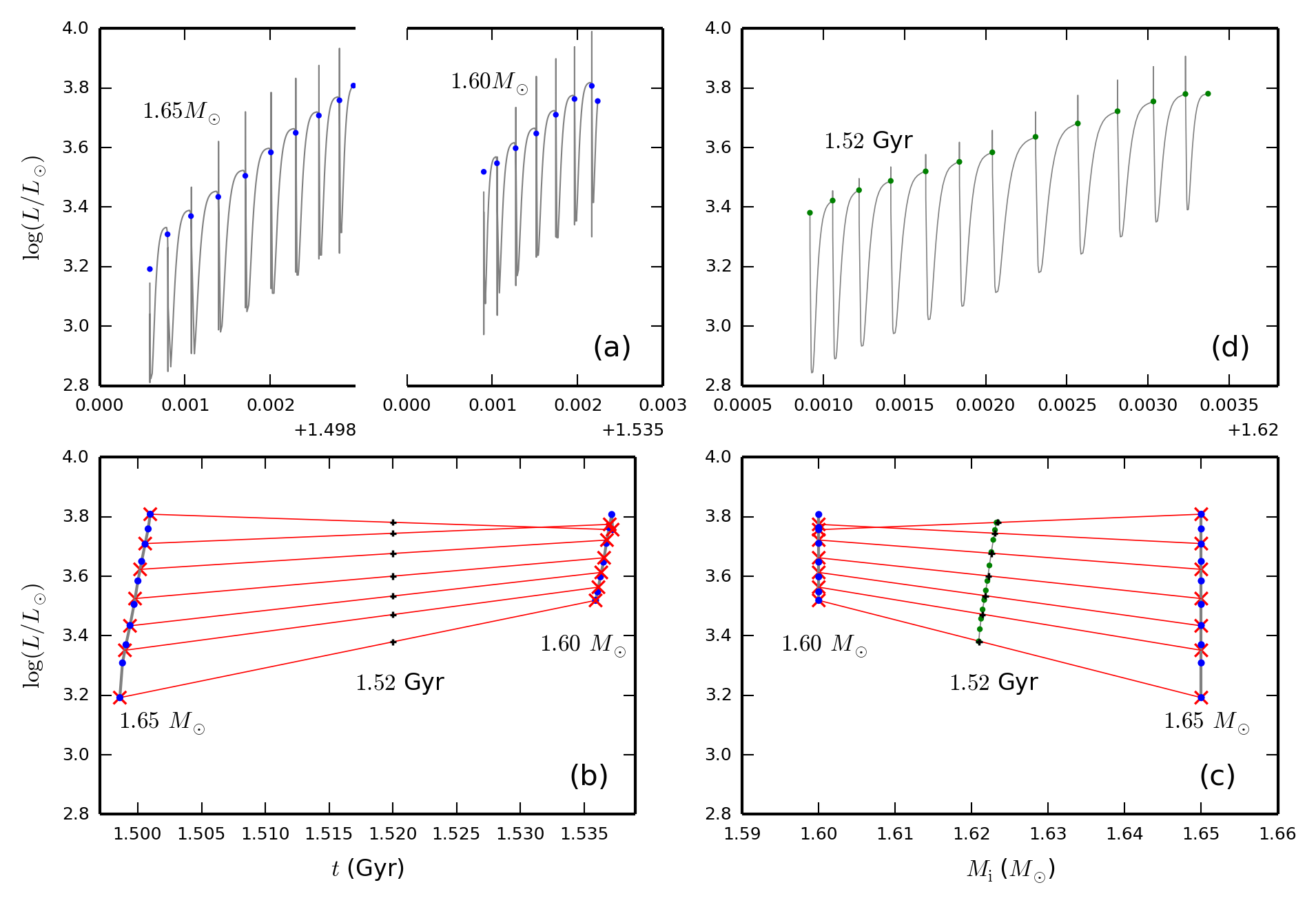}
\caption{The sequence of operations involved in building a TP-AGB section of the isochrones, limited to the stellar property $\log L$ only. Left panels refer to evolutionary tracks, while right panels refer to the derived isochrones. The split panel (a) shows a couple of \colibri\ tracks of same metallicity and close in initial mass (with $\mini=1.65$ and 1.60~\Msun, from left to right), in the absolute age $t$ versus $\log L$ plot. The continuous gray lines are the detailed TP-AGB evolution produced by \colibri, whereas the blue dots mark the pre-flash quiescent stages, or $L_\mathrm{q}$. Panel (b) shows the same two tracks in a much simplified way, containing just the pre-flash quiescent stages. They draw simple sequences in the $t$ versus $\log L$ plane, which can be easily split into a series of equivalent sections, as illustrated in this case by splitting each track into 6 sections of equal $\Delta t$ (delimited by the red crosses). Linear interpolations among these couple of equivalent points allow us to derive the $\log L$ for any arbitrary mass or age located between these tracks, as illustrated here by producing a series of intermediate points at a fixed age of 1.52~Gyr (black plus signs). Panel (c) shows the same situation as panel (b), but now in the $\Mini$ versus $\log L$ plane, showing the initial mass \Mini\ of every point produced at an age of 1.52~Gyr; they define a TP-AGB isochrone section containing only quiescent stages. The plot shows a few additional green points, which represent the somewhat denser grid of interpolated quiescent stages that is produced, by default, by our code. Panel (d) shows the final 1.52~Gyr isochrone in detail, after the detailed TPCs are re-introduced between the quiescent TP-AGB stages. }
\label{fig:tpcinterp}
\end{figure*}

\begin{figure*}
\includegraphics[width=\textwidth]{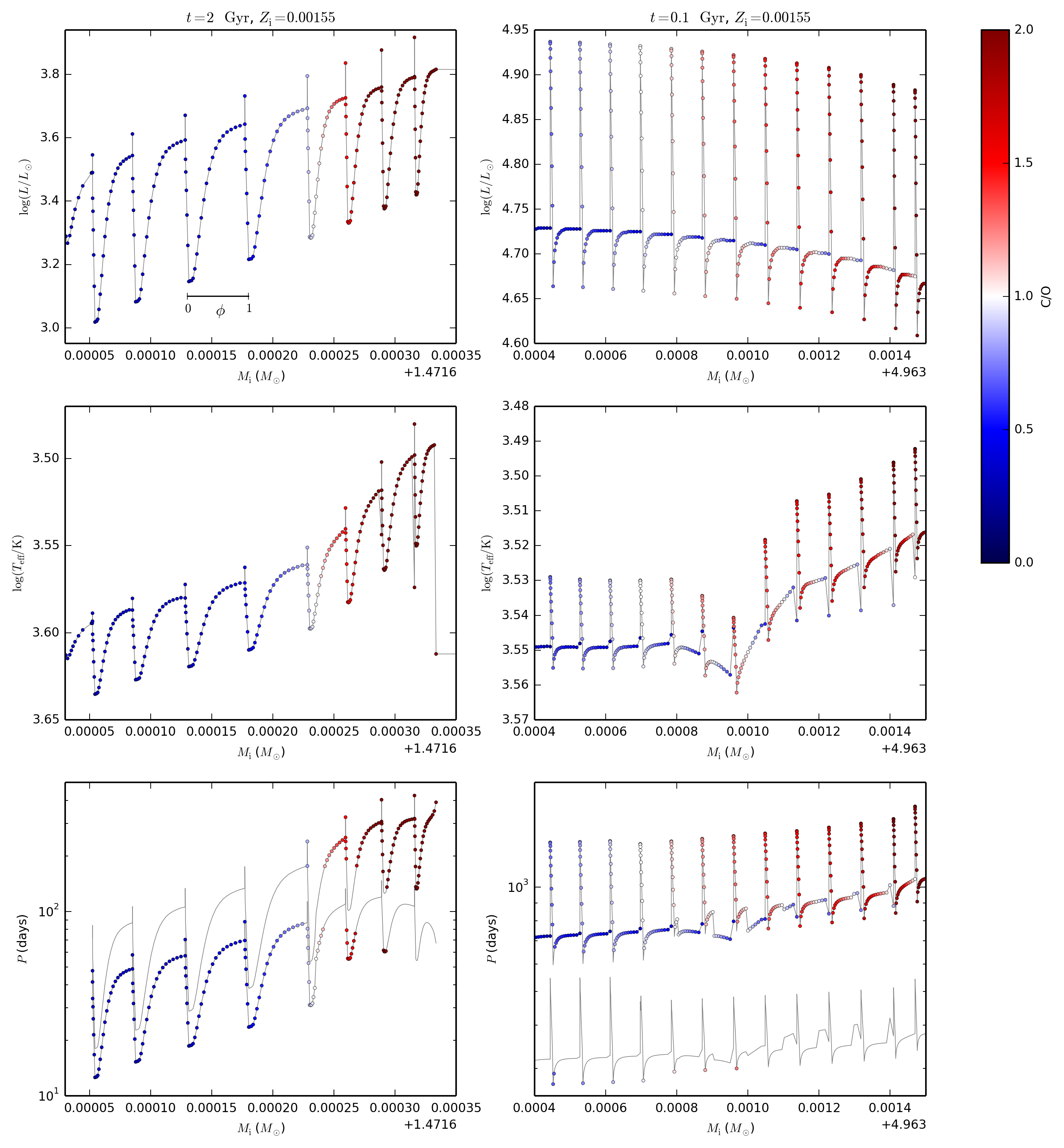}
\caption{Examples of isochrones, zooming on a few of their TPCs in the $\log L$, $\log\Teff$, and period vs.\ \mini\ plots (from top to bottom). In the top and middle panels, the dots along the isochrones are those derived for $\nintp=25$. The inset axis illustrates the variation of phase $\phi$ along one of the inserted TPCs. The left panels show the complete TP-AGB section of an intermediate-age isochrone, where third dredge-up is operating and the \co\ ratio steadily increases with \mini. This isochrone describes the sequence from M to S to C-type stars (i.e.\ from $\co\!<\!1$ to $\co\!\sim\!1$ to $\co\!>\!1$, cf.\ the color scale). The right panels shows a much younger isochrone, where HBB is operating in addition to the third dredge up. In this case HBB becomes gradually weaker for stars of higher \mini, since they correspond to stars in more advanced stages of the TP-AGB, in which there is a significant reduction of the envelope (and total) masses. This situation results into the sequence from M to S to C types to occur in reverse order {\em inside individual TPCs}. Finally, the bottom panels show the variation of LPV periods along these isochrone sections, for the fundamental mode and first overtone modes (the grey lines with longer and shorter periods, respectively). The colored dots in this case signal the expected dominant period.}
\label{fig:tps}
\end{figure*}

Being the numbers of TPC different, it is practically impossible to obtain TPC-looking features from the direct interpolation between adjacent TP-AGB evolutionary tracks. A typical situation is illustrated in panel (a) of Fig.~\ref{fig:tpcinterp}, where \colibri\ tracks of masses 1.60 and 1.65~\Msun, and $\Zi=0.001$, are plotted in absolute age versus \logL. Although the tracks look similar in slope and span comparable age and \logL\ intervals, the 1.65~\Msun\ track has 9 TPCs (out of which, just 8 are really complete), whereas the 1.60~\Msun\ has 7 TPCs.

To solve this problem, we apply a scheme that was devised during the first implementation of TP-AGB tracks in the population synthesis code \trilegal\ \citep{girardi07}: First, the complex TP-AGB tracks from \colibri\ are converted into simplified ones containing just the pre-flash quiescent stages\footnote{Pre-flash quiescent stages are those in which the H-burning shell provides most of the stellar luminosity, immediately prior the ignition of the He-shell in a flash; see \citet{wagenhuber98} for a precise definition.}. A couple of such simplified tracks are illustrated in Fig.~\ref{fig:tpcinterp}b. They look much smoother than the original tracks, and it is easy to make interpolations along them to derive versions containing more or less evolutionary points. In the illustrative case of Fig.~\ref{fig:tpcinterp}b, we have generated simplified tracks containing 6 equally-lasting TP-AGB sections. This number of sub-intervals can be set at any arbitrary value; in general, however, we set it at being equal to the maximum number of TPCs found in the mass interval being taken into consideration.

Quantities between these simplified tracks are interpolated using the same EEP scheme used for the pre-TP-AGB tracks; since they draw smooth lines in the Hertzsprung-Russell (HR) diagram, the derived isochrone sections will also contain a smooth and well-behaved sequence of quiescent pre-flash stages, as can be seen in the interpolated isochrone section of Fig.~\ref{fig:tpcinterp}c.

In addition to the $L$, other quantities characterizing the quiescent stages are stored and interpolated while creating the TP-AGB isochrone sections: they include \Teff, the core mass \Mcore, and the surface chemical composition. These quantities suffice to describe the complete luminosity profile  $L(\phi)/L_{\rm q}$ as a function of the phase during the TPC,  $\phi$, normalized to the pre-flash luminosity maximum $L_{\rm q}$, using the formulas provided by \citet{wagenhuber96} and \citet{wagenhuber98} \citep[just as done in \colibri\ by][]{marigo13}. Therefore, we first produce the isochrone TP-AGB sections containing just the quiescent luminosities (Fig.~\ref{fig:tpcinterp}c), and then we insert a number of points \nintp\ between them, following the detailed TPC luminosity evolution. This last step is illustrated in Fig.~\ref{fig:tpcinterp}d. In this example we have adopted $\nintp=15$, which suffices to recover the main details of the TPCs. Thanks to our choices for the number of age intervals in the simplified TP-AGB tracks, the number of quiescent points on every isochrone (and hence of TPCs being inserted) turns out similar to the number found in the track with a mass equal to the turn-off one.

At this point, we have isochrones with detailed $L$ variations along the TPCs, $\Delta\log L$, as further illustrated in the upper panels of Fig.~\ref{fig:tps} for two particular isochrone sections: one intermediate-age isochrone at around the point in which third dredge-up drives the transition between O-rich and C-rich phases, and a young isochrone at around the point in which mass-loss weakens the effect of HBB allowing the same transition to occur. The remarkable difference in the TPC shape between these cases is caused mainly by their very different core and envelope masses.

\subsection{Thermal pulse cycle $\Teff$ variations}

Having the $L$ variations being re-constructed, the next problem is to insert the detailed \Teff\ variations in the isochrones. For most TPCs in the original set of \colibri\ tracks, these \Teff\ variations closely follow the $L$ variations, as the star goes up and down along its Hayashi line in the HR diagram. Indeed, we verify that every single TPC develop along lines of nearly constant slope $\Delta\log\Teff/\Delta\log L$.
We recall that this behavior is derived from detailed envelope integrations in \colibri, which take into account the opacity variations deriving from changes in atomic and molecular concentrations.  $\Delta\log\Teff/\Delta\log L$ is generally well-behaved along the O-rich section of the TP-AGB tracks (depending primarily on the initial metallicity), but its value changes abruptly from about $-0.12$ to $-0.16$ whenever a transition between O-rich and C-rich star occurs -- following the large variations in molecular opacities in the stellar atmospheres \citep{marigo02}. In addition to this general behavior, we have the changes in \Teff\ due to substantial mass loss occurring at the last TPCs, in which while the luminosity increases mildly, \Teff\ decreases dramatically due to large mass loss which reduces the envelope mass. Thus, the star moves between Hayashi lines of different masses. In this case, it is not appropriate to model the \Teff\ variations simply with $\Delta\log\Teff/\Delta\log L$ slopes. Therefore, in each TPC we fit \Teff\ as a function of $L$ and the envelope mass $M_{\rm env}$, simultaneously, as $\log \Teff = a_0 + a_1\log L + a_2 M_{\rm env}$. The envelope mass $M_{\rm env}$ is computed through the current mass $M$ and the core mass \Mcore, as $M_{\rm env} = M - \Mcore$. Besides the quantities at the quiescent stages, we also store the linear fitting parameters within each TPC of the original \colibri\ set of tracks, and interpolate them while going from the quiescent tracks to the isochrones. These interpolated fitting parameters allow us to easily convert the TPC $L$ and $M_{\rm env}$ variations into $\Teff$ variations. 
 
\Teff\ variations computed in this way are illustrated in the middle panels of Fig.~\ref{fig:tps}. They clearly show the main change in the behaviour of $\Teff$ that occurs at the O- to C-rich transition. In the case of the younger isochrone (right panels), this event is further complicated by the strong mass loss that is associated to every one of its TP-AGB points.
 
Other stellar properties are interpolated and/or re-constructed along the TPCs in the isochrones, in particular the pulsation periods and chemical abundances. These are discussed later in Sects.~\ref{sec:tpcperiods} to \ref{sec:otherq}.

\subsection{Further computational details}

Although the scheme described and illustrate above is quite general, a couple of additional details help us to produce isochrone TP-AGB sections more closely resembling those of the original \colibri\ tracks. The first one is to adopt $\log\Mini$ as the independent variable in all interpolations involving mass, instead of $\Mini$; this choice produces smoother isochrones whenever the grids contain tracks widely spaced in \Mini, but in reality it makes little difference for the present, dense grids of tracks. The second detail is that, whenever two adjacent TP-AGB tracks contain a C-star phase, those tracks are split into two equivalent sections. In other words, we place an additional EEP at the O- to C-rich transition, hence imposing that the interpolations between tracks occur largely internally to their main O-rich and C-rich sections. In this way we avoid that the main change in $\Delta\log\Teff/\Delta\log L$ slope between C- and O-rich stars, causes artificial features in the shape of the derived isochrones\footnote{Later on in the evolution, as the star loses most of its envelope and starts to heat in its way towards the planetary nebulae stage, the $\Delta \log\Teff/\Delta\log L$ slopes change appreciably again, becoming even positive. This late change is also naturally present in the final section of the isochrones.}.
 
Moreover, we note that the TPC points are \textit{not} inserted at evenly spaced intervals of \mini\ along the isochrones. Instead, they are more closely spaced at the beginning of the TPCs, which present the largest variations in $\log L$ (see Fig.~\ref{fig:tps}). 

\subsection{Spectral libraries for cool giants}

The \parsec\ isochrones are transformed into absolute magnitudes via a series of bolometric correction (BC) tables which are suitably interpolated in the \logg\ versus \logte\ plane, and taking into account the surface chemical abundances, i.e.
\begin{equation}
M_\lambda = M_\mathrm{bol} - \mathrm{BC}({X_i},\logg,\logte) \,\,\,,
\label{eq:absmags}
\end{equation}
where $M_\mathrm{bol}=-2.5\,\log(L/L_\odot)+4.77$.

The formalism to derive the BC tables starting from libraries of synthetic spectra is fully described in \citet{girardi02,girardi10}.

For most stars, the ${X_i}$ interpolation of Eq.~\ref{eq:absmags} means simply a linear interpolation in the variable $\mh$ -- hence a 3D linear interpolation in $\logte\times\logg\times\mh$ space. The main exception to this rule regards exactly the most luminoust TP-AGB stars, which are treated according to a different scheme, reflecting the more extended sets of atmospheric models built to describe them:

\begin{figure*}
\includegraphics[width=1\textwidth]{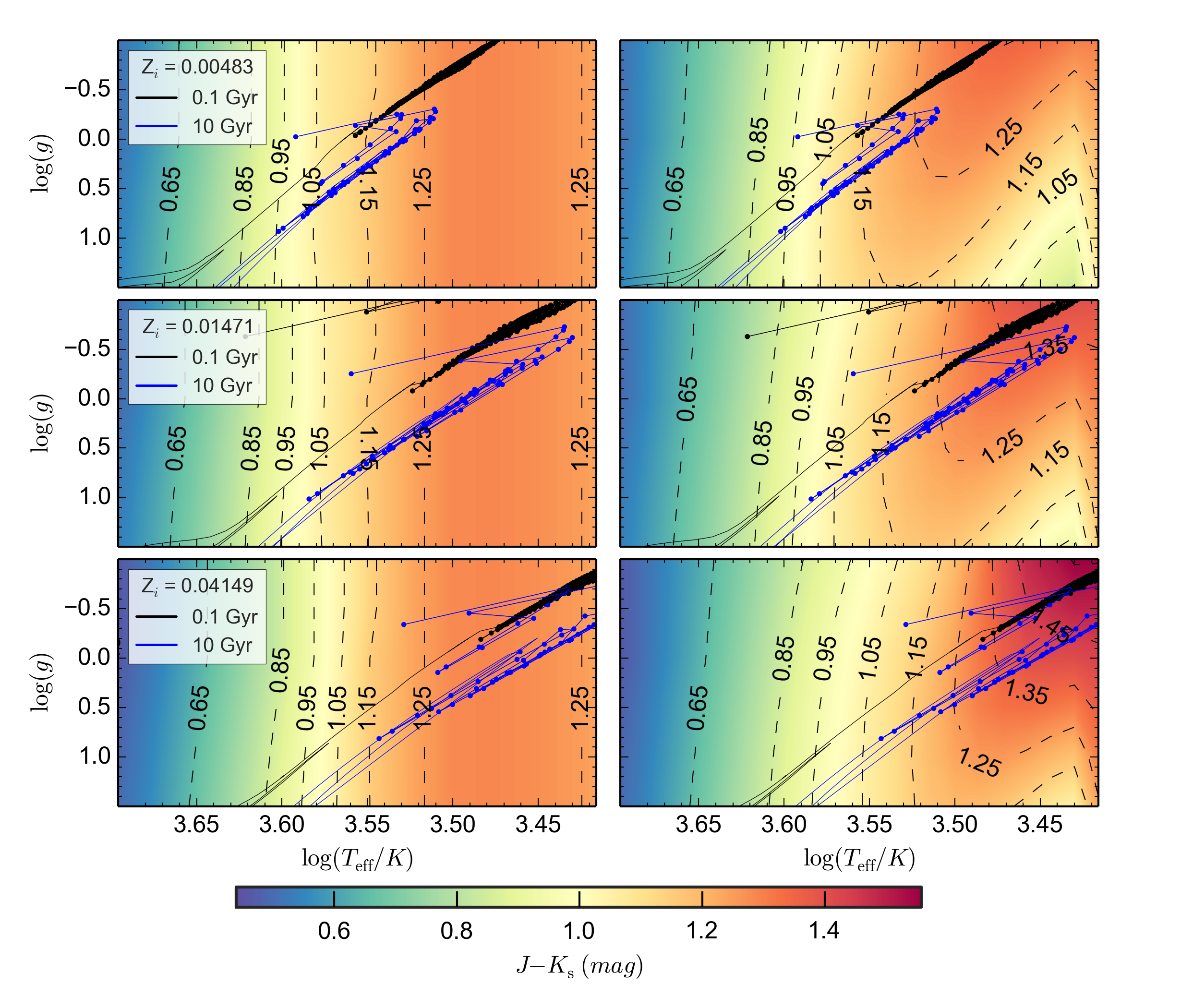}
\caption{Map of the 2MASS \jks\ colours attributed to O-rich giants of metallicities $\mh=-0.5$, 0, and $+0.5$ (top, middle and bottom panels, respectively), as a function of \logte\ and \logg, for two cases: the previous spectral library used in \citet{marigo08} (left panels), and the present one which incorporates the \citet{aringer16} results (right panels). A couple of isochrones with ages of 0.1 and 10 Gyr and initial $Z_\mathrm{i}=0.00483$ (top panels), $Z_\mathrm{i}=0.01471$ (middle panels) and $Z_\mathrm{i}=0.04149$ (bottom panels) are shown for comparison, displaying in particular the TP-AGB part -- recognizable by the zig-zag during thermal-pulse cycles, at $\logg\lesssim0.5$. Only the O-rich section of these isochrones is plotted. We note in particular that with the present prescriptions (right panels) the colours of the coolest giants depend on \logg\ and \mh, while with the former prescriptions (left panels), only the dependence with \Teff\ was being considered.}
\label{fig:bc}
\end{figure*}

\subsubsection{Carbon-rich giants}

C-rich giants (with a surface carbon-to-oxygen ratio larger than 1, $\mathrm{C/O}>1$) have their atmospheres rich in carbon-bearing molecules like CN, C$_2$, C$_3$, HCN and C$_2$H$_2$, presenting spectra significantly different than O-rich giants of similar parameters. For this reason, a separate database of C-rich spectra is needed. The \citet{marigo08} isochrones distributed since 2009 have used either the \citet{loidl01} or the \citet{aringer09} databases of synthetic C-rich spectra. For the present work, this library has been partially replaced: for solar metallicities (i.e. for the same content of heavy metals, excluding C and O) we implement a new set of spectral calculations provided by \citet{aringer16}, based on the latest version of the \texttt{COMARCS} hydrostatic atmosphere models. This set comprises over 950 models with parameters in the ranges $4000<\Teff/\mathrm{K}<2500$, $2<\logg<-1$, $\mathrm{C/O}=[1.01, 1.05, 1.1, 1.4, 2.0]$, and masses of $M/\Msun=[1, 1.5, 2, 3]$. Most of the computed spectra, however, correspond to stars of low masses (1 and 2 \Msun), with $\logg<0$ and $\Teff<3400$~K, which are the intervals of parameters more relevant to describe the TP-AGB C stars observed in nearby galaxies. 

The interpolation of bolometric corrections inside the C-rich grid follows the procedure outlined in \citet{aringer09}: in short, the spectral properties are interpolated inside the multidimensional grid with primary parameters being $Z$, $\Teff$, \logg, and C/O. Then, a small correction due to the spectral variations with mass (which, for a fixed \Teff\ and \logg, corresponds to adopting an atmospheric model with a different sphericity) is applied. 

In practice, the latest update in the C-star library affects all isochrones containing C-type stars at metallicities larger than $\mh=-0.5$. For these stars, the bolometric corrections will be based not only on the new models, but also benefit from the better interpolation inside a richer grid of models. In this context it should be noted that the new database also covers objects with C/O ratios as low as 1.01, while the smallest value in \citet{aringer09} was 1.05.

\subsubsection{Oxygen-rich giants}

Regarding the ``normal'' O-rich stars (with $\mathrm{C/O}<1$), in previous releases the spectral database was either the ATLAS9-based from \citet{castelli03} or the PHOENIX from \citet[][see \citealt{chen14} for details]{allard12}, replaced by \citet{fluks94} for cool M giants. 
\citet{aringer16} has recently published a huge database of O-rich spectra derived from the \texttt{COMARCS} code, especially suited to describe the spectra of cool giants. Following the indications from this latter paper, we now adopt a smooth transition between the BCs derived from the previous databases, and those from \citet{aringer16}, in the \Teff\ interval between 4000 and 5000~K ($4.6<\log(\Teff/\mathrm{K})<4.7$).  This interval roughly correspond to the CHeB phase (including the low-mass red clump) in solar-metallicity isochrones, and affects only the brightest RGB, TP-AGB and RSG stars of metallicities $\mh\lesssim-1$. In most optical and infrared passbands, such a transition will appear completely smooth. Some artifacts might appear at UV wavelengths, where the different spectral libraries are provided with a very different sampling in $\lambda$. However, having precise UV photometry of cool stars is somewhat unlikely, so that this aspect might be less of a problem.

One of the most important consequences of adopting the \citet{aringer16} spectral library is that of having a more physically-sound description of how the molecular lines vary as a function of stellar parameters. Especially important is the behaviour of water lines which heavily determine the near-infrared colours of O-rich giants; depending on their detailed behaviour, near-infrared colours such as \jh\ and \jks\ may not increase monotonically with decreasing \Teff, as could be naively expected. This effect can be appreciated in Fig.~\ref{fig:bc}, which illustrates the changes we have in the 2MASS $\jks$ colours of metal-rich O-rich giants, as a function of \logte\ and \logg, for metallicities going from a third of solar to three times solar ($\mh=-0.5$, 0 and $+0.5$), and comparing present prescriptions based on \citet{aringer16} with those adopted in the previous \citet{marigo08} isochrones. As can be appreciated, \jks\ tends to reach an almost-constant value for the coolest giants, with $\Teff\lesssim3000$~K. The exact color of this ``saturation'', and the \Teff\ at which it occurs, depends heavily on the total metallicity -- which is crucial in determining the efficiency of water formation -- and mildly also on the chosen linelist \citep[see][for details]{aringer16}. Moreover, the \jks\ colors depend also on the exact $\logg$ reached by the giants; for the more compact giants (generally corresponding to the stars of lower initial mass stars found in older isochrones) there is even the possibility that the $\jks\times\Teff$ relation reverses, with the coolest giants becoming slightly bluer than the $\Teff\sim3000$~K ones. Fig.~\ref{fig:bc} shows that this complex behaviour is present in the new isochrones when the \citet{aringer16} spectral library is adopted. 

All isochrones with O-rich sequences cool enough to enter in the $\Teff\lesssim3000$~K range will be affected by this change in the spectral library of O-rich giants. This regards especially isochrones of solar and super-solar metallicity, like those illustrated in the middle and bottom panels of Fig.~\ref{fig:bc}. In isochrones of metallicities $\mh\lesssim-0.5$~dex (e.g.~those in the upper panels of Fig.~\ref{fig:bc}), O-rich sections are generally hotter than this \Teff\ limit. Moreover, at smaller metallicities a larger fraction of the TP-AGB appears as C-rich (see Sect.~\ref{sec:bulk} below).

It is also interesting to note that the spectral library from \citet{aringer16}, by including C/O ratios up to 0.97, approaches the region of parameters covered by S-type stars, in which the scarcity of free C and O and the enrichment of s-process elements give place for the appearance of molecular features of species such as ZrO and YO. More extended computations, fully covering the transition between C-, S- and M-type spectra, will be provided in \citeinprep{Aringer et al.}

One can also notice in Fig.~\ref{fig:bc} that the present isochrones reach smaller \logg\ and cooler \Teff\ than the $\logg<-1$, $\Teff>2600$~K limits of the \citet{aringer16} spectral library -- especially at young ages (TP-AGB stars with HBB) or for very old and metal-rich isochrones. There is no easy solution for this problem, since the hydrostatic model atmospheres from \texttt{COMARCS} are both (a) hard to converge for stars of low \Teff, and (b) not realistic given the rise of high-amplitude pulsation and of huge convective cells at the stellar photosphere. Notice however that such very cool stars are rare, and that their observed properties are severely affected by circumstellar dust (cf. next subsection), hence little reflecting the detailed photospheric properties. However, for practical purposes these stars also need to be attributed a photospheric magnitude and colour; we do so by simply extrapolating the bolometric correction tables linearly in the $\logg\times\log\Teff$ plane. 

\subsection{Circumstellar dust}
\label{sec:dust}

On top of the photospheric spectra, our models include the effect of light reprocessing by circumstellar dust in the extended envelopes of mass-losing stars for which we computed the radiative transfer (RT). The novelty in the present isochrones is the inclusion of a self-consistent treatment of dust growth, fully calculated as a function of the input stellar parameters, i.e. luminosity, actual mass, effective temperature, mass-loss and elemental abundances in the atmosphere, as described in \citet{Nanni13, Nanni14}.
Our dust growth description provides the dust mixture as a function of the stellar parameters as well as the optical depth at $\lambda=1\mu{\rm m}$.
The dust code is coupled with a RT one \citep[\texttt{MoD} by][]{Groenewegen12}, based on \texttt{DUSTY} \citep{Ivezic97}.
The most recent update of our dust growth model \citep{Nanni16} regards the production of carbon dust in circumstellar envelopes of C-stars, which are particularly relevant for the interpretation of the colors of the so-called extreme-AGB stars observed in infrared surveys of nearby galaxies \citep[e.g.][]{boyer11}. 
Since in Circumstellar Envelopes of C-stars the bulk of the dust produced is mainly composed by amorphous carbon, the dust temperature at the inner boundary of the dust zone is assumed to be the one of carbon dust, even if we also include silicon carbide (SiC) and iron dust in our calculations. For the same reason, for M-stars the dust temperature at the inner boundary of the dust zone is assumed to be the one of the first silicate dust condensed (either pyroxene or olivine). For M-stars the calculations also take into account the formation of Al$_2$O$_3$, quartz (SiO$_2$), periclase (MgO) and iron.

Therefore, the approach adopted allows us to compute bolometric corrections self-consistently as a function of stellar parameters, and provides the corresponding change in the broad-band absolute magnitudes and colors. A large grid of such models was computed, covering 2 values of masses (0.8 and 2~\Msun), 5 of mass loss (from $10^{-7}$ to $10^{-5}$~\Msun\,${\rm yr}^{-1}$ at logarithmic spaced intervals), 3 luminosities ($\log(L/L_\odot)=3.25$, 3.75 and 4.25), five \Teff\ (from 2600 to 3400~K), 3 metallicities ($Z=0.001$, 0.004, and 0.008), and 6 values of carbon excess for C stars (with $8.0<\log(n_{\rm C}-n_{\rm O}) -\log(n_{\rm H})+12<8.8$, where $n_{\rm C}$, $n_{\rm O}$, and $n_{\rm H}$ are the surface number fractions of carbon, oxygen, and hydrogen respectively).
The bolometric correction applied to every star in the isochrones is determined via a multi-dimensional interpolation in this grid. For the O-rich stars, the models adopt the same parameters and optical data as in \citet{Nanni13,Nanni14}.
For the C-rich stars, two grids are presently provided. One is computed using the \citet{rouleau91} set of optical data with typical size of dust grains $\sim 0.1$ $\mu$m, while the other is computed by employing \citet{Jaeger98}'s dataset produced at a temperature of $400$~K with grains of size $\sim 0.06$ $\mu$m. As detailed in \citet{Nanni16}, among several possible choices of optical data and parameters, these two are the ones that best reproduce the infrared colors of C stars in the SMC. However, larger deviations for the \citet{rouleau91} optical data set are expected for the reddest colors, for which smaller grains ($\sim 0.06$ $\mu$m) are better in reproducing the data.

The treatment of dust and radiative transfer adopted in this new release of isochrones represents a remarkable improvement with respect to the previous prescriptions assumed in \citet{marigo08}.  There, the interpolation was performed for tables of spectra pre-computed for few dust mixtures \citep{bressan98, groenewegen06}, since a full model for the growth and destruction of dust was still not developed. 

\subsection{Interstellar dust}

In addition, the effect of heterochromatic interstellar dust extinction is considered in the isochrones as in \citet{girardi10}. This means applying, to the points along the isochrones, the extinction coefficients which are a function not only of the passband being considered, but also of the stellar $\Teff$, of the total extinction in the $V$ band, $A_V$, and of the ratio between selective and absolute extinction $R_V$, as defined in \citet{cardelli89} and \citet{odonnell94}'s generalized extinction curve.

\subsection{Long period variability}

\begin{table}[]
\centering
\caption{Coefficients for the periods in Eq.~\ref{eq:periods}.}
\label{tab:periods}
\begin{tabular}{cc|cccc}\hline
 m & T & $a$ & $b$ & $c$ & $d$ \\ \hline
 0 & O & $-$0.683& 1.874 & $-$0.109 & $-$1.957 \\
 0 & C & $-$0.757& 2.018 & $-$0.121 & $-$2.282 \\
 1 & O & $-$0.585& 1.620 &    0.083 & $-$1.641 \\
 1 & C & $-$0.499& 1.515 &    0.107 & $-$1.406 
\end{tabular}
\end{table}
 
As demonstrated by extensive microlensing surveys of the Magellanic Clouds and Milky Way Bulge, TP-AGB stars are frequently observed as long-period-variables (LPV), pulsating in at least one of several pulsation modes between the fourth overtone and the fundamental (Miras) mode \citep{lattanzio04, wood15}. LPV properties were introduced in isochrones for the first time in \citet{marigo08}, using a series of fitting relations derived from \citet{fox82}, \citet{wood83} and \citet{ostlie86} to describe periods in the fundamental and first overtone modes ($P_0$ and $P_1$ respectively), as well as luminosity at which the dominant period changes between these modes. These relations are presently being revised with the aid of extensive calculations of pulsation models for a wide grid of input parameters, computed with linear non-adiabatic radial oscillation code described in \citet{wood14}. For the present work, we update the fitting formula for the predicted $P_0$ and $P_1$. They are of the form:
\begin{eqnarray}
\log(P_\mathrm{m}^\mathrm{T}/\mathrm{days}) & = & a\log(M/\Msun) + b\log(R/R_\odot) + \nonumber \\
    & & c\log(M_\mathrm{env}/M) + d
\label{eq:periods}
\end{eqnarray}
where m stands for the mode (0/1 for fundamental/first overtone) and T for the chemical type (O- or C-rich). The fitting coefficients are presented in Table~\ref{tab:periods}. These relations where derived from models along \colibri\ evolutionary tracks covering the $Z_\mathrm{i}$ interval from $0.002$ to 0.017, and for masses between 0.8 to 2.6~\Msun. The fitting relations provide periods accurate to within a few per cent. Further updates and a thorough discussion will be presented in \citeinprep{Trabucchi et al.}. The likely transition luminosity between the first overtone to fundamental mode was computed in the same way as in \citet{marigo08}, based on \citet{ostlie86}. 

Importantly, $P_0$ and $P_1$ pulsation periods are computed along the isochrones but the periodic changes in the photometry are not taken into account. Therefore, the photometric properties we provide should be regarded as \textit{mean properties} over the LPV pulsation periods, rather than instantaneous values.

\section{Results}
\label{results}

Here, we give just a quick overview of the main novelties in the present isochrones:

\subsection{Surface chemical abundances along isochrones}

\begin{figure*}
\includegraphics[width=\textwidth]{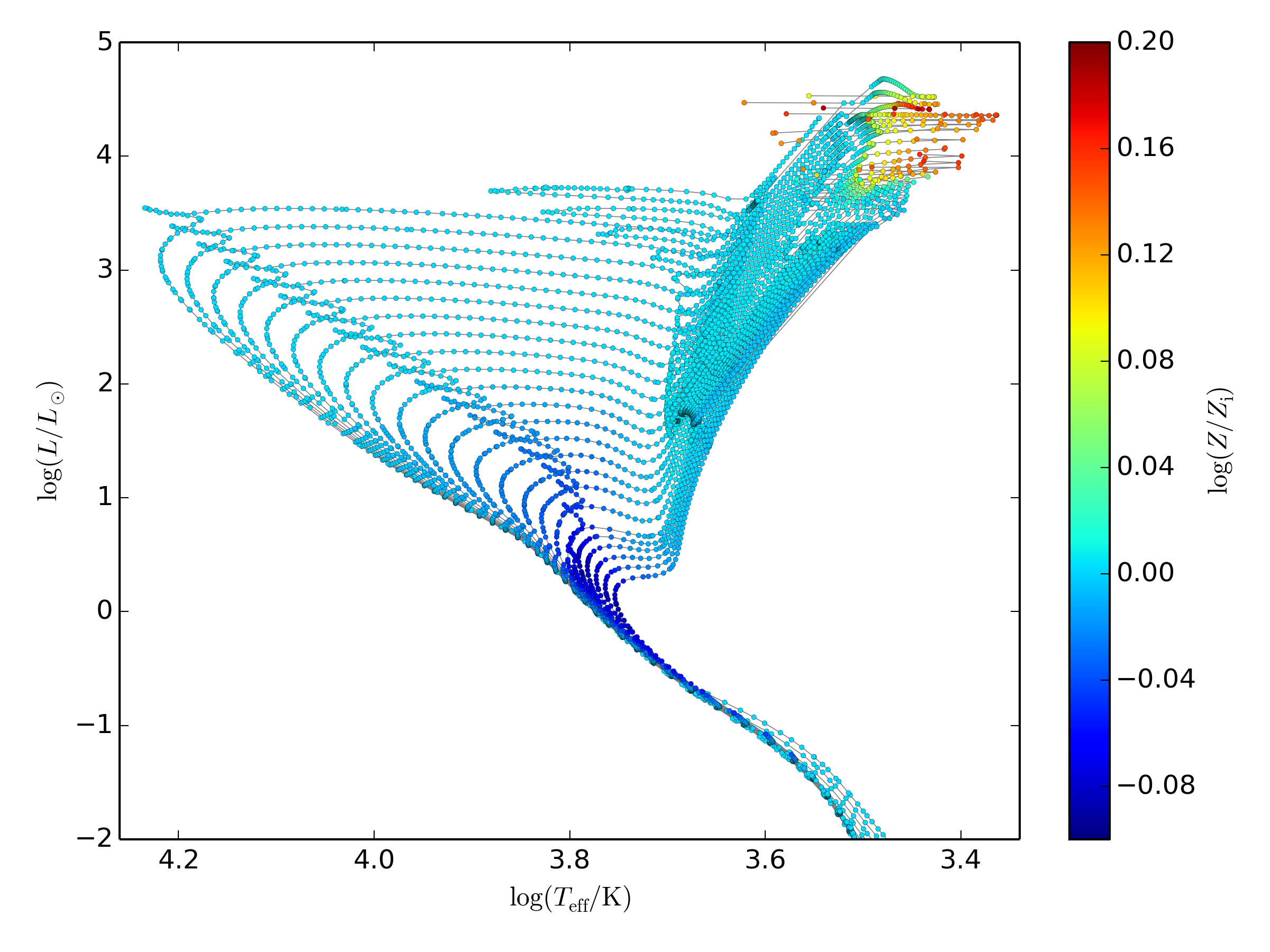}
\caption{Changes in the current surface metal content $Z$ across the HR diagram, for a set of isochrones of initial $Z_\mathrm{i}=0.01471$, with ages varying from $\log(t/\mathrm{yr})=7.8$ to 10.1 at steps of 0.1~dex, \textit{and for $\nintp=0$}.}
\label{fig:abund}
\end{figure*}

As already commented, with this isochrone release we start providing detailed surface abundances $\{X_i\}$ -- presently consisting of the mass fractions $X$, $Y$, $Z$, $X_{\rm C}$, $X_{\rm N}$, and $X_{\rm O}$ -- {\em along} the isochrones. These values are all interpolated linearly (always using the EEPs as a reference) from those given along the evolutionary tracks. Their values change as a result of microscopic diffusion (especially during the main sequence of low-mass stars), of dredge-up events, and of HBB at the most massive TP-AGB stars. Fig.~\ref{fig:abund} illustrates the regions of the HR diagram which are most affected by these changes, for a set of given initial metallicity, by showing the changes in the total metal content by mass, $Z$. The main changes in $Z$ occur at the main sequence of low-mass stars due to microscopic diffusion (the purple area), and at the TP-AGB due to third dredge-up events (the yellow-to-red sections at the top). In addition, the isochrones also contain the variations in $Z$ due to the second dredge-up in intermediate-mass stars, and the almost-imperceptible variations along the RGB due to protons locked up in CNO nuclei, which appear after the first dredge-up.

\begin{figure}
\includegraphics[width=\columnwidth]{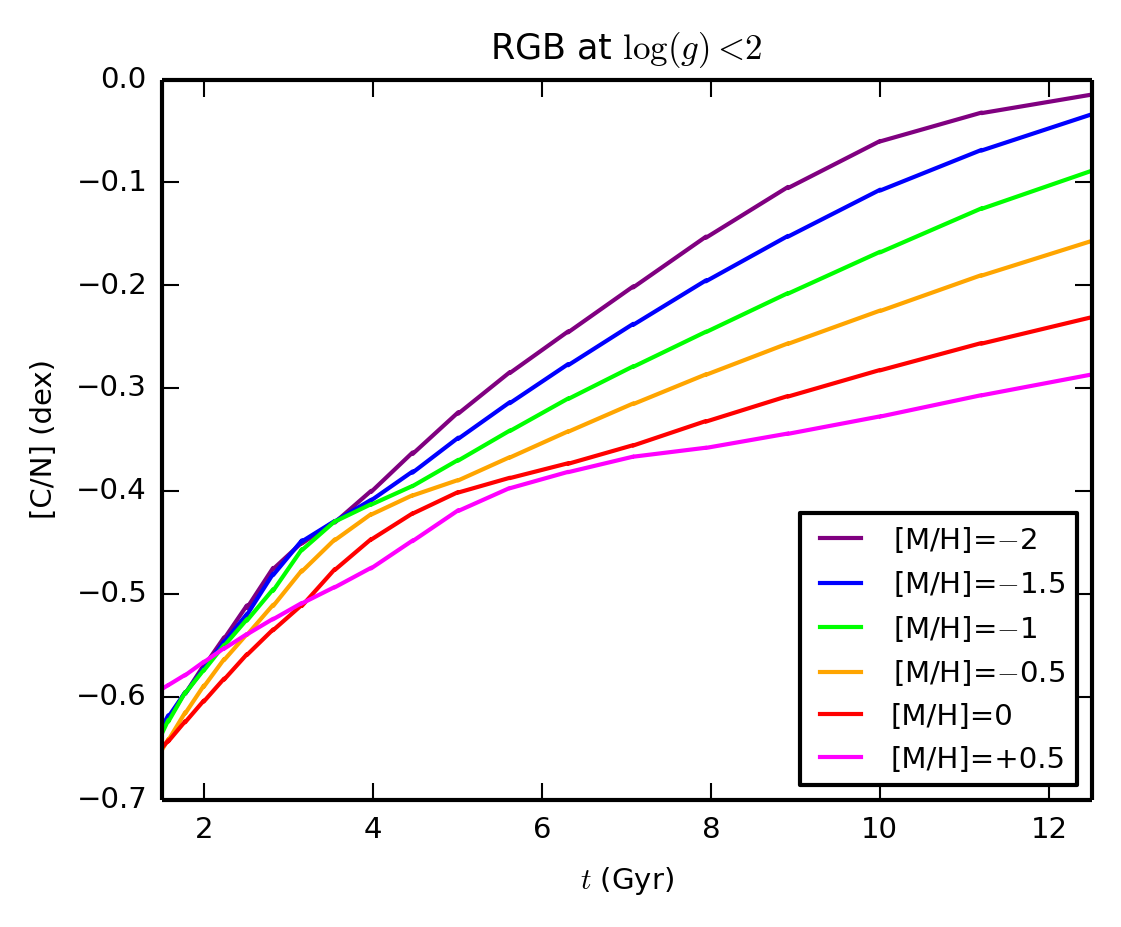}
\caption{The predicted variation of [C/N] with age  in RGB stars of a few selected metallicities. }
\label{fig:cn}
\end{figure}

Figure~\ref{fig:cn} illustrates one of the many possible applications of the new abundance information: its left panel presents the expected correlation between the number ratio of C and N as compared to the Sun, $[\mathrm{C/N}]$, and the age of RGB stars. This relation results from the first dredge-up in low-mass stars, as discussed in \citet{salaris15}. As can be seen, it presents a clear dependence on the initial metallicity, which is also well understood\footnote{The somewhat different slope for the youngest isochrones of $\mh=0.5$~dex, likely results from its very high initial helium content, $Y_{\mathrm i}=0.32$.}. Such a relation is nowadays, directly or indirectly, being used to constrain ages of field MW stars \citep[see][]{masseron15, ness15, martig16}. Having C and N abundances tabulated in grids of isochrones means that the observed samples of field RGB stars can be more easily modelled via population synthesis approaches, giving us the possibility to check for possible biases or contaminants in the observations.

Despite the potential importance of the $[\mathrm{C/N}]$ versus age relation, its empirical verification has been somewhat limited so far. \citet{salaris15} present a comparison between the $[\mathrm{C/N}]$ results from BaSTI models and data for 5 open clusters plus a point representative of the halo field; although the general trend with age is confirmed, the comparison clearly suffers from the few observational points, from the large abundance errors, and likely from the heterogeneity of the cluster data.

\subsection{Surface chemical abundances along TPCs}
\label{sec:tpcchemistry}

Another main novelty in the present isochrones is the presence of detailed information regarding the stellar changes -- in $L$, \Teff, and $\{X_i\}$ -- {\em along} TPCs. We note that \citet{marigo08} isochrones with TPC variations have already been distributed upon request in the past, but are now provided as an option ready-to-use in the new web interface. 

Figure~\ref{fig:tps} shows how the TPCs appear in the \Mini\ vs.\ \logL\ diagram of two isochrones, illustrating (1) the notable differences in the shape of TPCs between low- and intermediate-mass stars \citep[see also][]{boothroyd88, vassiliadis93}, and (2) the sizeable variations in surface chemical compositions along these isochrone sections. 

This aspect of providing chemical abundance changes along TPCs deserves a more detailed description. Let us start with the simplest case: low-mass stars in which only 3DU is operating. In the original evolutionary tracks, the surface chemical composition of these stars change soon after the He-flash that initiates the TPC (i.e.\ at a phase $\phi=0$), and stay constant along the entire TPC until its end, just prior to the next He-flash (at $\phi=1$). In the isochrones, our interpolation algorithm transforms this step-by-step change in the surface abundances into a gradual change, as can be appreciated in the left panels of Fig.~\ref{fig:tps}. This feature is intentional: indeed, our isochrones aim at explaining the expected surface abundances in large populations of TP-AGB stars, in a statistical way, and not the behaviour that happens in every individual TP-AGB star as it evolves. In fact, the important point in this case is that the general trend of varying abundances with the main stellar parameters -- mass, age, luminosity, etc. -- is preserved by our interpolations.

In the case of the more massive intermediate-mass stars, in addition to 3DU we have that HBB changes the abundances (and sometimes also the spectral type) {\em along every single TPC}. This feature also needs to be reproduced in the isochrones, since it causes a correlation between the chemical composition and other stellar properties; for instance, it causes stars found in their low-luminosity dip phases of a TPC to be more likely C stars than stars found in brighter, quiescent phases, even if other stellar properties (age, core mass, initial mass and metallicity, etc.) are similar. Therefore, we also include this effect in the isochrones, by distributing the in-cycle abundances changes caused by HBB, $\Delta(X_i)$ (which values are also interpolated among TP-AGB tracks of all masses and metallicities), along every single TPC re-created in the isochrones. The result can be appreciated in the right panels of Fig.~\ref{fig:tps}, which show an isochrone section where \co\ transits several times between values lower and higher than 1.

\subsection{\nintp\ and the general appearance in color-magnitude diagrams}
\label{sec:nintp}

\begin{figure*}
\includegraphics[width=\textwidth]{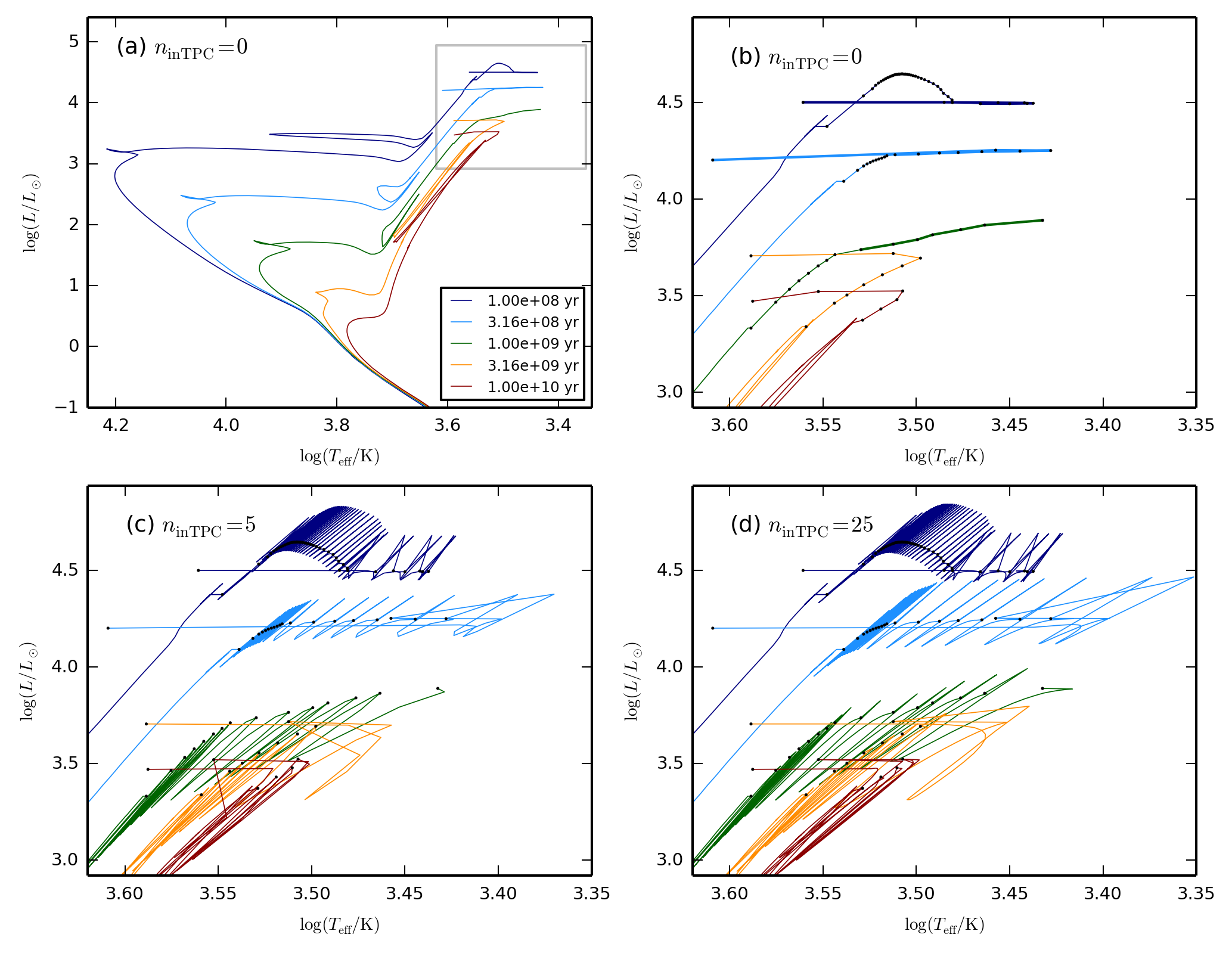}
\caption{Examples of isochrones in the HR diagram, for a fixed metallicity ($\mh=-0.4$~dex, or $Z_\mathrm{i}=0.006$) and 5 different ages at equally-spaced $\Delta\log t=0.5$~dex intervals. Panel (a) shows a large section of the HRD for isochrones computed with $\nintp=0$ -- that is, without the details of TPCs. Panels (b) to (d) zoom on the TP-AGB region of pabel (a), showing a sequence of increasing $\nintp$. The small black dots mark the position of quiescent points along the TPCs. On panel (b) only, the C-rich section of the isochrones is drawn with a heavier line. }
\label{fig:isoc}
\end{figure*}

\begin{figure*}
\includegraphics[width=\textwidth]{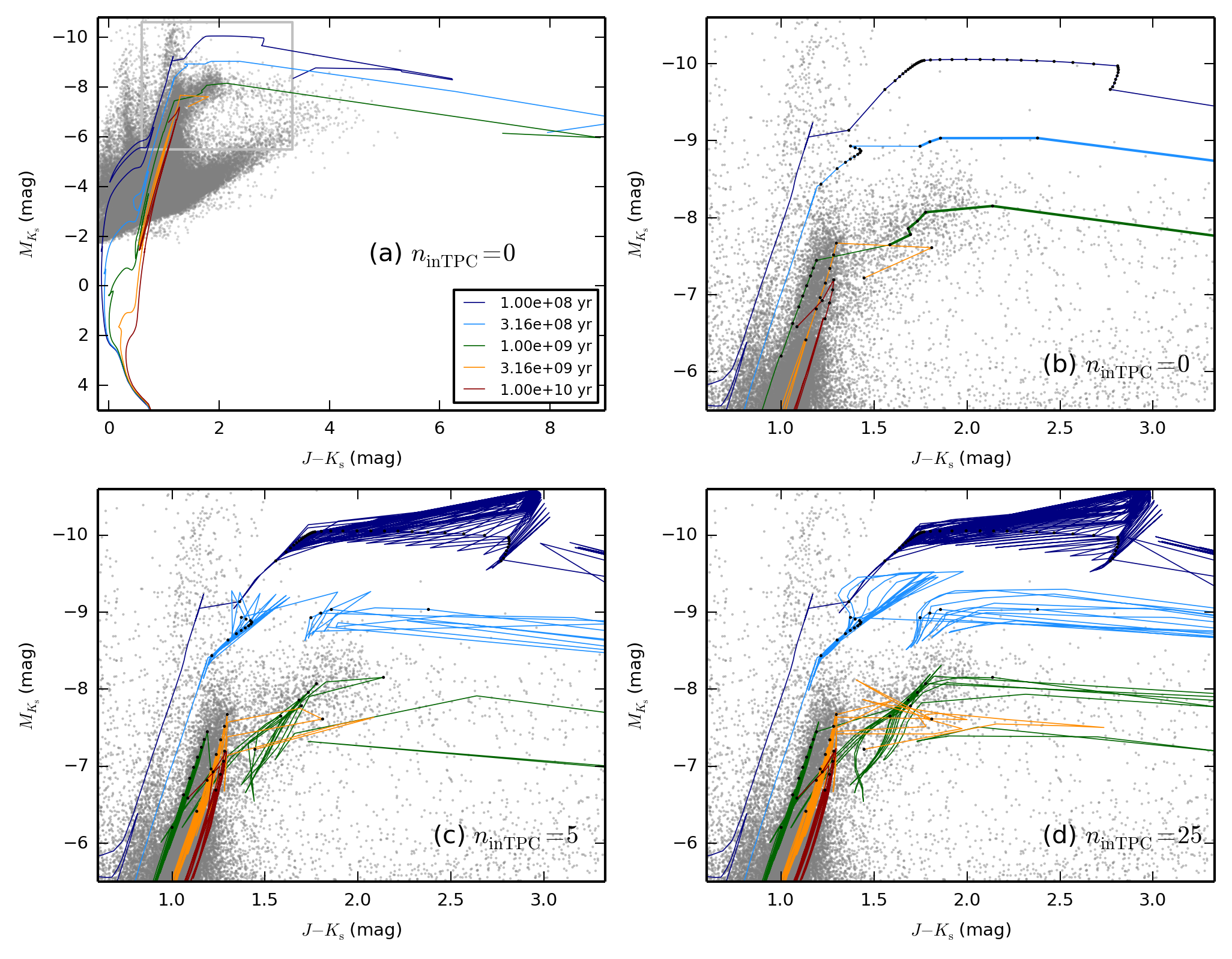}
\caption{The same as in Fig.~\ref{fig:isoc}, but for the classical CMD of 2MASS photometry, $M_{K_\mathrm{s}}$ vs. $\jks$. Grey points in the background are the 2MASS point sources \citep{2mass} within 1$^\circ$ of the LMC center as determined by \citet{vandermarel01}; they are shifted by a true distance modulus of 18.5~mag, whereas extinction is ignored.}
\label{fig:isoc2}
\end{figure*}

Figure~\ref{fig:isoc} illustrates a set of isochrones of initial metallicity $\mh=-0.4$~dex ($\Zi=0.006$) and covering a wide range of ages, in the HR diagram. The different panels show how the TP-AGB sections of the isochrones appear for different values of \nintp. The initial case of $\nintp=0$ corresponds to smooth isochrones, containing only the quiescent points of TPCs -- similarly to those distributed by \citet{marigo08}. The isochrones with $\nintp=5$ contain already a fair amount of detail about TPCs, including its most important feature: the relatively long-lived low-luminosity dips, which are described by a set of three points in every TPC. In isochrones with $\nintp=25$ the $L$-$\Teff$ variations are described in greater detail; their low-luminosity dips appear as smooth sequences, and they present slightly more detailed high-luminosity peaks. Therefore one can set the following general guidelines as regards the \nintp\ value: in simple applications that just require an overall description of the main loci of TP-AGB stars (for instance, looking at variations in the C-rich sequences with age, as in panel b of Fig.~\ref{fig:isoc}), one can use $\nintp=0$. But for a fair description of the main {\em strips} occupied by TP-AGB stars in different diagrams, one needs to consider the well-populated low-luminosity dips, for which a \nintp\ of at least 5 is recommended. For more detailed population studies aimed at accurately reproducing star counts, \nintp\ can be increased to values as high as $>20$.

This point is illustrated further in Figure~\ref{fig:isoc2}, which presents the same isochrones in the popular $\ks$ vs.\ $\jks$ diagram of 2MASS photometry. The isochrones with $\nintp>0$ present CMD sequences significantly broadened due to the TPCs, as expected. Remarkably, along the extreme-AGB region the $J-\ks$ color variations inside TPCs become so wide, that increasing $\nintp$ to 25 does not suffice to transform the isochrone features into smooth lines. Therefore, the most suitable value of $\nintp$ may also depend on other additional factors like the wavelength of the observations to be interpreted, their photometric depth and typical photometric errors, etc. For all these reasons, \nintp\ is kept as a free parameter in our codes and web interfaces. 

That said, it is interesting to note how O- and C-rich stars are predicted to distribute in different regions of the HRDs and CMDs of Figs.~\ref{fig:isoc} and \ref{fig:isoc2}. First of all, there is the well-known fact that a C-rich phase is only found in isochrones younger that a few Gyr. In addition, in the HRD, there is an evident trend of C stars being found at smaller \Teff\ than O-rich stars; this effect is caused by the sudden increase in Rosseland mean opacities as C/O overcomes the value of 1 in stellar atmospheres \citep{marigo02,marigo13}. In the $J\ks$ CMD, the bulk of C stars separates even more from the O-rich ones: most O-rich stars tend to distribute along a broad diagonal sequence that extends from $[\ks,\jks]=[-6, 1.0]$ to $[-10, 1.4]$. C stars instead crearly draw a separate sequence departing from the O-rich one at $[-7,1.3]$ and reaching $[1.8, -9]$. The latter is caused, in addition to the \Teff\ changes, by the different spectral features of C and M stars, including those caused by circumstellar dust. 

All these features became well-known since the release of near-infrared photometry for the Magellanic Clouds \citep[e.g.][]{cioni00, nikolaev00}. A very preliminary comparison with 2MASS data for the inner LMC is presented in Fig.~\ref{fig:isoc2}, just evincing a few facts: First, the TP-AGB sections of the younger isochrones (with ages 1 to $3.16\times10^8$~yr) have very few counterparts in 2MASS data; this is likely caused by the fact that these TP-AGB sections, being derived from TP-AGB models with small lifetimes, actually span very narrow intervals in initial mass. Isochrones with ages larger than 1~Gyr instead have clear correspondence in the overall distribution of observed points. The most prominent features in these diagrams, in addition to the tip of the RGB (TRGB) at $-$6.5, are:
\begin{itemize} 
\item the well-populated and diagonal sequence of O-rich AGB stars extending from the TRGB up to $M_{K\mathrm{s}}$ and at $\jks<1.3$, which is well matched by the isochrones with $t>1$~Gyr; 
\item an upward, less-populated extension of this O-rich sequence up to $-9.5$, which upper part is well matched by our $3.16\times10^8$-yr isochrone; 
\item the red tail of C-rich stars extending up to $\jks\sim2$, which is well described by the C-rich section of our 1-Gyr isochrone; 
\item and finally the scarsely-populated and extended ``extreme AGB'' region, defined by $J-\ks>2$, which slope is well matched by our isochrones with ages $3.16\times10^8$ and $10^9$~yr.  
\end{itemize}
This qualitative comparison with 2MASS does not consider the expected occupation probability of every isochrone section -- which can be easily derived from the isochrone tables -- as well as other fundamental aspects such as the IMF, and history of star formation and chemical enrichment in the LMC. A detailed quantitative comparison with the Magellanic Clouds' data will follow in separate papers.

As a cautionary note, we note that the appearance of the extreme-AGB sequences in the CMD, is still affected by limitations in our present grids of RT models: since they do not cover the entire range of parameters found in the tracks, models with the most extreme $L$, C/O, $\dot{M}$ and \Teff\ have their properties badly determined. This applies especially to the stars at the top-right boundary of the ``extreme-AGB'' region, corresponding just to the youngest isochrones in Fig.~\ref{fig:isoc}, with an age of $10^8$~yr. This problem, other than temporary (since RT models are being expanded) is not a major one if we consider that just a minority of the AGB stars observed in nearby galaxies are expected to be affected.

\subsection{Period changes along TPCs}
\label{sec:tpcperiods}
 
In the isochrone TP-AGB sections, pulsation periods change from point to point according to equation~\ref{eq:periods}, thanks to the large changes in $M$, $L$, \Teff, and $\{X_i\}$, both between isochrones and inside TPCs of single isochrones. Although the pattern of these changes can be rather complicated, a few general trends can be drawn here, with the help of Fig.~\ref{fig:tps}: At low luminosities and relatively high \Teff, the dominant pulsation mode is the first-overtone one, of shorter period. Most isochrone TP-AGB sections will start with this dominant mode. Later along the isochrone, a transition to a dominant fundamental mode (longer period) can occur. As illustrated in the bottom panels of Fig.~\ref{fig:tps}, the transition happens along a few TPCs. In the isochrones of intermediate-age populations, the transition from first-overtone to fundamental mode often coincides with the main transition from O- to C-rich phases, where a marked reduction of the \Teff\ takes place \citep{marigo02,marigo07}; whenever this happens, however, there are always small C-rich sections expected to appear in the first-overtone, as there are O-rich sections expected in the fundamental mode.

For younger isochrones where the TP-AGB develops at high luminosities and cool {\Teff}s, larger sections of the isochrones appear with a dominant fundamental mode. This includes most of the stars undergoing HBB, with their characteristic high $L$ and low C/O ratio.

In reality, the situation is more complex with stars appearing in multiple pulsation modes and in higher-order overtones. More complete modeling of these sequences is underway (Trabucchi et al., in prep.), and will be taken into account in subsequent releases of the isochrones. For the moment, it suffices to mention that our predictions for the fundamental mode pulsation seem the more robust, since in this case both the entrance (via the transition luminosity given by  \citealt{ostlie86} relation) and the final exit (usually via mass loss) from the FM sequence are modelled consistently. The same cannot be said of the first overtone, for which there is not yet a physically-based prescription defining the moment stars enter on this pulsation mode.

\subsection{Other quantities added in (or cleared from) the isochrone tables}
\label{sec:otherq}

The present isochrone tables present a few other useful quantities derived from the already-mentioned ones, for instance, the surface gravity $g$ and the radius $R$, which are simply derived by means of its physical definition and from the Stefan-Boltzmann law, respectively. There is also a label to identify the main evolutionary stages -- pre-main sequence, main sequence, subgiant branch, RGB, CHeB, early-AGB, TP-AGB -- as well as the last point on the TP-AGB. Also given are TP-AGB specific quantities like the C/O ratio (in number), the C excess in C-rich stars \citep[see][for its precise definition]{eriksson14}, the mass-loss rate, the optical depth of the circumstellar dust at 1$\mu$m, the LPV periods and pulsation modes.

With respect to \citet{marigo08}, we have also been more careful about the range of applicability of some tabulated stellar properties, and about their names. In particular:
\begin{itemize}
\item Pulsation periods and modes of long period variables are presented just for TP-AGB stars; otherwise they are set to null and $-1$, respectively. Indeed, while the relations used by to compute periods and modes may also apply to bright RGB and RSG stars, they have not been tested for those stars. Moreover, they would not provide reliable periods for fainter variables like classical Cepheids and RR Lyrae. 
\item Presently, we keep track of the H-exhausted core masses only along the TP-AGB phase, where this quantity is useful for determining the properties of TPCs. Therefore, core masses are named as $M_{\rm core,TP}$. They are set to null for any evolutionary phase previous to the TP-AGB.
\item Likewise, the dust optical depth, $\tau(1\mu{\rm m})$, is set to null for all evolutionary phases previous to the TP-AGB, even if their mass losses are significant (as close to the tip of the RGB phase). The reason for this choice is that the mechanism for dust formation may act differently in other stars, and may even be inefficient in RGB stars \citep{gail98}. Moreover, present-day claims of detection of circumstellar dust shells around RGB stars, as well as their interpretations in terms of mass loss, are still either controversial \citep[see][and references therein]{boyer10, momany12} or depend on observations of very small samples of nearby stars \citep{groen12}.
\end{itemize}
All these points may be revised in future releases of these isochrones, as the theoretical prescriptions are replaced and/or expanded, and other useful quantities and abundance information are added to the tables.
   
\subsection{TP-AGB bulk properties}
\label{sec:bulk}

A notable difference of the new TP-AGB tracks computed with the \colibri\ code, compared to those of \citet{marigo07}, is related to the duration of the TP-AGB phase. These differences are illustrated in detail in the recent review by \citet{marigo15} and in \citet{rosenfield16}. A particular prediction is that the numbers of C stars make an ``island'' in the metallicity vs.\ age plane, that is: their formation is strongly favoured at ages peaking at $\sim2$~Gyr and at metallicities $-1<\mh<0$, while they nearly disappear at all other extremes of the age-metallicity plane. This prediction is well described in figure 7 of \citet{marigo15}, which shows a map of the TP-AGB lifetime of the C stars expected at every age and metallicity.

\begin{figure*}
\includegraphics[width=\textwidth]{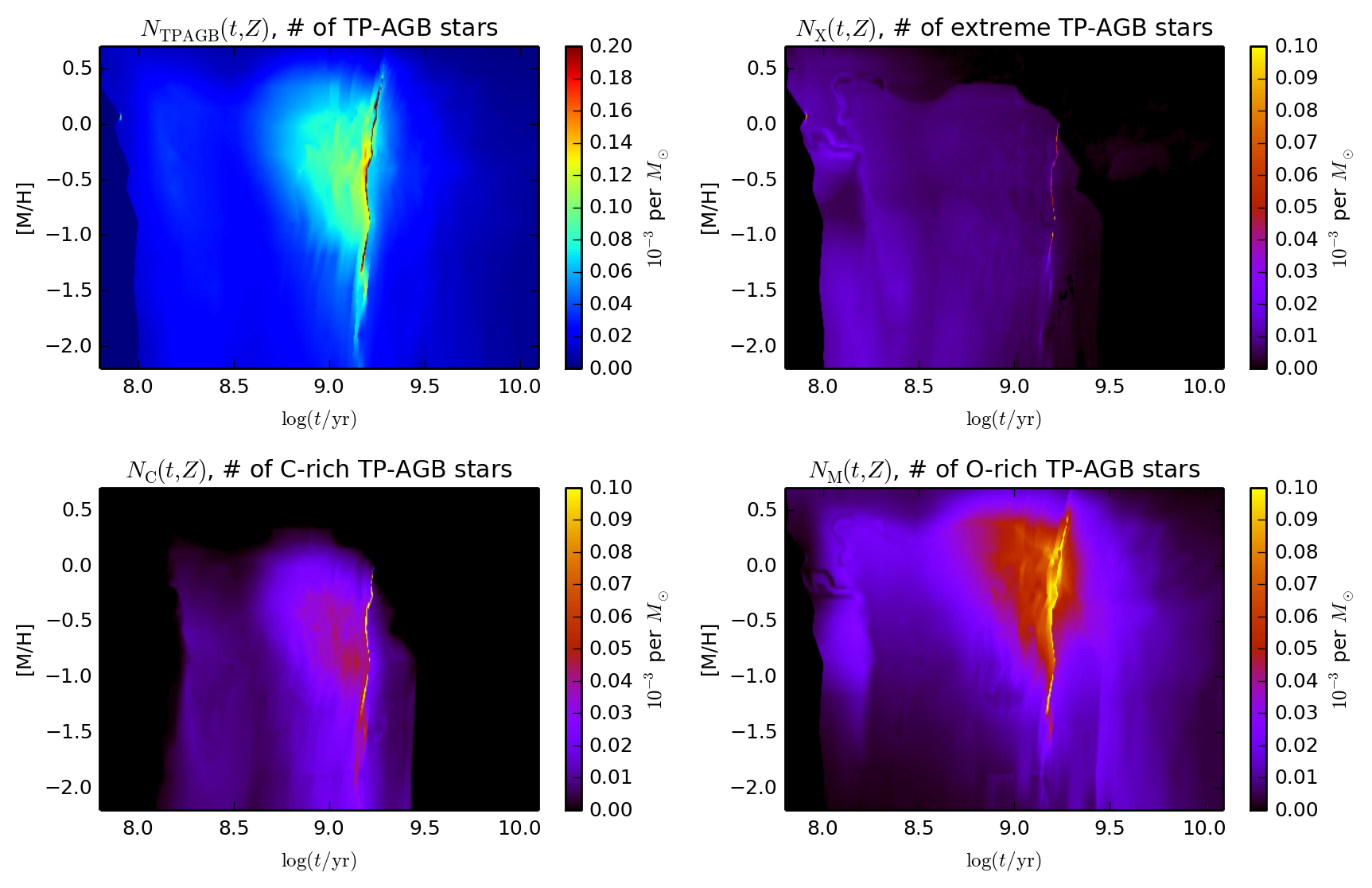}
\caption{Map of the the expected numbers of TP-AGB stars as a function of the initial metallicity and age, per unit mass of formed stars in a stellar population. They have been computed using the \citet{kroupa02} IMF, and adopting very simple criteria for the identification of C, M, and extreme TP-AGB stars. The first panel also presents, on a distint color scale, the total numbers of TP-AGB stars. See the text for more explanation.}
\label{fig:nstar}
\end{figure*}

With the detailed isochrones, we can check these predictions in more detail. First of all, we can eliminate the approximation (widely adopted since \citealt{frogel90}, and implicitly assumed in \citealt{marigo15}) that the number of TP-AGB stars in a stellar population is proportional to the lifetime of the TP-AGB tracks with the same turn-off mass. Indeed, \citet{girardi13} demonstrated that this approximation badly fails at ages close to 1.6~Gyr when a rapid change in core He-burning lifetimes cause a marked spike in the production rate of TP-AGB stars \citep[the ``AGB-boosting'' effect, see also][]{girardi98}. Moreover, another difference with respect to previous works regards the youngest isochrones containing TP-AGB stars with HBB, for which we can now separate C- and M-type stars along individual TPCs (Fig.~\ref{fig:tps}), and correctly compute their mean properties.  

Thus, Fig.~\ref{fig:nstar} shows the expected number of different kinds of TP-AGB stars per unit mass of stars formed in a stellar population, as a function of age and metallicity. These quantities derive from the simple integral
\begin{equation}
N_\mathrm{TPAGB}^{j}(t,Z) = \int_{(t,Z)}^j \phi(\mini) d\mini
\end{equation}
where $\phi(\mini)$ is the initial mass function (IMF) by number, normalised to a unit total mass, i.e.
\begin{equation}
\int_0^{\infty} \mini \phi(\mini) d\mini=1\Msun \,\,\,,
\end{equation}
and the integral is performed over the isochrone sections $j$ that satisfy the desired criterion. In practice, this calculation is greatly facilitated by the isochrone tables being produced as sequences of increasing initial mass \Mini, and including tags for the evolutionary stage which clearly indicate isochrone sections where the TP-AGB phase is either starting or terminating. This allows us to easily replace the integral by a summation that only considers the pairs of {\em subsequent} points which satisfy the condition $j$.

For the sake of illustration, we consider four simple criteria in Fig.~\ref{fig:nstar}. The first one  illustrates the total numbers of TP-AGB stars ($N_{\rm TP}$). Then, three criteria were designed to separate these TP-AGB in broad categories: those of extreme TP-AGB stars ($N_{\rm X}$, with $\jks>2$~mag) which generally have mass loss rates in excess of $|\dot{M}|>10^{-6}$~\Msun/yr and are optically-obscured by their circumstellar dust shells, plus those of optically-visible C and M-type TP-AGB stars ($N_{\rm C}$ and $N_{\rm M}$, with $\jks<2$~mag and ${\rm C/O}>1$ and ${\rm C/O}<1$, respectively).  Of course, these definitions do not correspond precisely to those adopted in analyses of observed samples; however, similar estimates can be easily derived from the available isochrones, using alternative criteria based on the photometric or chemical properties of stars.

There are some remarkable aspects in the maps presented in Fig.~\ref{fig:nstar}:
\begin{enumerate}
\item First, it is evident that optically-visible C stars appear over a limited area of the metallicity--age plane, being absent at all ages $t>4$~Gyr. This limit is essentially caused by the low efficiency (or absence) of third dredge-up in stars of low masses. It also corresponds to a classical observational fact, namely the absence/scarsity of bright C stars (i.e.\ with luminosities clearly above the tip of the RGB) in the oldest clusters of the Magellanic Clouds, which has become clear since the compilation by \citet{frogel90}. Indeed, the oldest LMC cluster to clearly contain such C stars is NGC~1978, with 6 likely members and an age of about 2~Gyr \citep{mucciarelli07}. In the SMC, the oldest case is Kron~3, with 3 likely members and an age of $\approx6.5$~Gyr \citep{glatt08}\footnote{The SMC clusters NGC~339 and NGC~121, with ages of about 6.5 and 11.5~Gyr respectively \citep[see][]{glatt08,glatt08b}, have 1 C star each. These however are underluminous objects and might correspond to the result of binary evolution, which is not being considered here.}.
\item C stars are also absent in young metal-rich populations, and become progressively more frequent in young populations of metallicity $\mh<-0.5$~dex. This absence at young ages is caused mainly by HBB in the more massive TP-AGB stars preventing a transition to the C-rich stage while the cluster is still optically-visible. This is also observed in the form of a paucity of C stars in young Magellanic Cloud clusters, as first noticed by \citet{frogel90}. Although the data is severely affected by the low number statistics and the uncertainties in the field contamination\footnote{The most impressive exception is the single C star observed in the core of the $\approx 50$-Myr old cluster NGC~1850, which, in the absence of independent indications for its membership, has usually been ascribed to the underlying LMC field.}, it is compatible with no visible C stars being observed in clusters younger than about 300~Myr \citep[turn-off masses larger than 3~\Msun, see figure 1 in][]{girardi07}.
\item Finally, C stars are absent from very metal-rich populations, with metallicities $\mh>0.4$~dex. This is caused both by more carbon dredge-up being needed to cause the transition to ${\rm C/O}>1$, and by the third dredge-up being weaker at higher metallicities. Conclusive evidence for this metallicity ceiling for the C-star formation has been recently found by \citet{boyer13} using Hubble Space Telescope medium-band observations of the inner M~31 disk. More specifically, they found that a population with mixed ages and a mean metallicity of $\mh\approx0.1\pm0.1$~dex has an extremely low number ratio between C and M stars,  $\mathrm{C/M}=(3.3^{+20}_{-0.1})\times 10^{-4}$.
\item A very unusual feature in all these $N^j(t,Z)$ plots is the nearly-vertical ``scar'' or ``lightning'' at ages close to $\log(t/{\rm yr})\simeq9.2$, which corresponds to the AGB-boosting effect described by \citet{girardi13}. The effect is determined by the rate at which stars leave evolutionary stages \textit{previous} to the TP-AGB, hence it appears indistinctly for all TP-AGB subsamples -- and nearly at the same age interval. It should be noticed however that the effect is extremely brief. It appears clearly in these plots thanks to their very fine age resolution (namely $\Delta\log t=0.005$~dex). \citet{girardi13} claims that this effect contributes to the maximum relative numbers of AGB stars observed in stars clusters such as NGC~419, NGC~1806, NGC~1846, NGC~1751, and NGC~1783, all with mean turn-off ages around 1.7~Gyr. \label{item:boosting}
\item With the exception of the AGB-boosting period, the TP-AGB production broadly peaks at ages between 1 and 2 Gyr. This feature is determined by the maximum TP-AGB lifetimes being reached for stars of initial masses in the range 1.5 to 2~\Msun\ \citep{marigo03, marigo07}. This feature also agrees with the maximum numbers of these stars, relative to the integrated cluster luminosity, being found in clusters with turn-off masses broadly located between 1.6 and 3~\Msun\ in the LMC \citep{girardi07}. 
\item Extreme AGB stars are more frequent at the youngest ages, where TP-AGB stars reach higher luminosities and have larger envelopes to lose through mass-loss. They become much rarer at ages exceeding 2~Gyr. It is hard to check this prediction quantitatively, since just a handful of Magellanic Cloud clusters contain extreme AGB stars, and usually just a few units per cluster \citep{tanabe04,vanloon05}. It is remarkable, however, that among the 30 stars with IR-excess likely belonging to clusters identified by \citet{vanloon05}, none is found in objects older than 3~Gyr (with turn-off masses smaller than $1.3$~\Msun), even if such clusters comprise $\sim\!10$\% of the total mass of their sample (see their table 9).
\end{enumerate}
Given the uncertainties both in the models and in the data -- the latter mostly regarding the field contamination, but also including uncertain age and metallicity estimates -- any comparison between them should be regarded just as a preliminary check of the mean trends.

\subsection{The AGB boosting period}
\label{sec:boost}

Regarding item \ref{item:boosting} above, we are completely aware that the AGB-boosting effect is in general absent from other published sets of isochrones in the literature. As discussed in \citet{girardi13}, the effect appears only if a fine mass resolution is adopted for the computation of the grids of evolutionary tracks, and if the interpolation between the tracks is properly dealt with, as a result of a quick decrease of CHeB lifetimes at ages of $\sim1.6$~Gyr. It is interesting to note that these conditions have been recently confirmed by \citet{choi16} with an independent set of evolutionary tracks. For works of evolutionary population synthesis of galaxies, this feature can be seen as an unnecessary complication, and it can be even intentionally suppressed \citep[see e.g.][]{dotter16}. However, as demonstrated in \citet{girardi13}, this period of intense TP-AGB production is an unavoidable consequence of the rearrangements in stellar lifetimes that appear as electron degeneracy develops in the H-exhausted cores of low-mass stars; it really corresponds to light emitted by stars, that {\em has to} be taken into account into the evolution of the bulk properties of stellar populations. Moreover, this feature is probably very important for the interpretation of the most classical TP-AGB calibrators, namely the intermediate-age star clusters in the Magellanic Clouds \citep{girardi13}, which in turn affect the calibration of the role of TP-AGB stars in population synthesis models themselves.

That said, it is a matter of fact that the AGB boosting is a quick event in the life of stellar populations, lasting for no longer that a few $10^8$~yr, so that it may appear or not depending on the exact values of ages chosen for producing isochrones around the 1.6~Gyr age interval. This point has been extenuously discussed and is well illustrated in figure 4 of \citet{girardi13}. 
Therefore, a careful selection of the age values and age bins might be necessary in order to have a fair account of the TP-AGB star counts and of their integrated light in population synthesis models of galaxies. 

As discussed in \citet{girardi13}, basing on isochrones derived from \parsec\ tracks computed with a mass spacing of $\Delta\Mi=0.01$~\Msun, the AGB boosting period turns out to last for $\Delta t_\mathrm{boost}\approx10^8$~yr (see their figure 4), yielding numbers of TP-AGB stars, $N^j(t)$, a factor $A_\mathrm{boost}\approx2$ larger than at immediately younger and older ages. The presently-released isochrones, instead, are derived from tracks computed with $\Delta\Mi=0.05$~\Msun\ in the mass range of interest here (at the boundary between RGB+HB and INT tracks in Fig.~\ref{fig:track_masses}, or $\Mini\sim1.75$~\Msun). This situation tends to produce AGB-boosting periods which are very short-lived (that is, $\Delta t_\mathrm{boost}$ is just a fraction of $\approx10^8$~yr) but producing a more pronounced spike in the TP-AGB numbers (hence a larger $A_\mathrm{boost}$ factor). The product of $\Delta t_\mathrm{boost} \times A_\mathrm{boost}$, however, seems to be roughly conserved, irrespective of the mass resolution of the tracks -- provided that this resolution is good enough to include the rise in the He-burning lifetimes that occurs at $\Mini\simeq1.75$~\Msun. We have verified that, given the present tracks with $\Delta\Mi=0.05$~\Msun\footnote{It is our intention to further improve this mass resolution in future releases of the \parsec--\colibri\ isochrones.}, the \textit{coarsest} age resolution necessary to detect the spike in $N^j(t)$ caused by the AGB-boosting period is of about $\Delta\log t\simeq0.005$~dex. Since this parameter can be set to any arbitrarily-small number, we recommend the galaxy-modeling community to use even finer resolutions in the vicinity of  $\log(t/{\rm yr})\simeq9.2$, so that the spike in the TP-AGB star numbers associated with the AGB-boosting effect can be detected and well sampled, and eventually distributed over the age bin considered in the modeling of galaxies.

\subsection{Comparison with other isochrones}
\label{sec:comparison}

\begin{figure*}
\includegraphics[width=\textwidth]{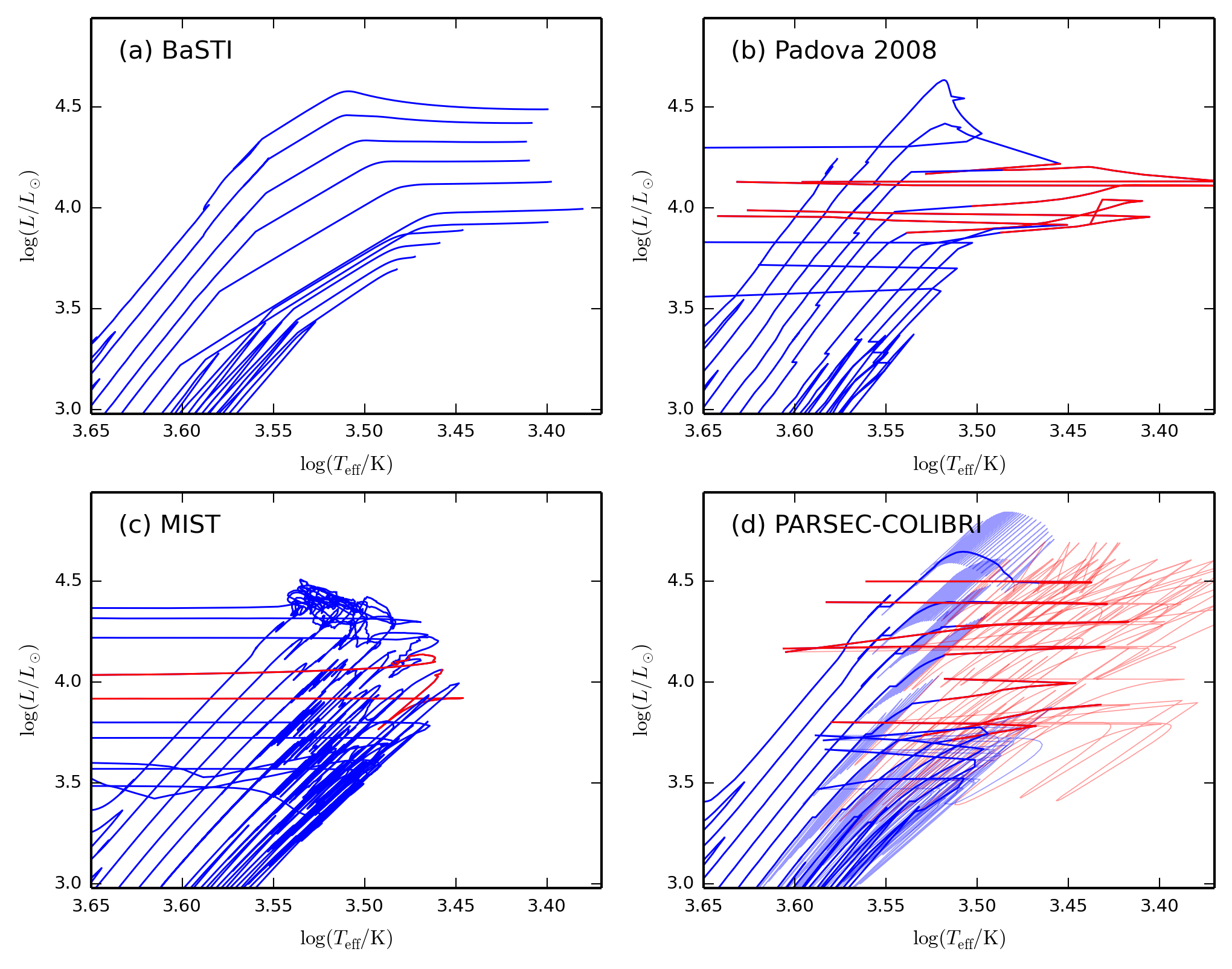}
\caption{
Comparison between available isochrones including the complete TP-AGB phase, in the HRD. All sets include convective core overshooting, and have a metallicity close to that of young LMC populations. In each panel, ages go from $\log(t/\mathrm{yr})=8$ to 10 at intervals of 0.2 dex, from top to bottom. O-rich sequences are marked in blue, C-rich ones in red. (a): BaSTI \citep{cordier07} isochrones for $\mh=-0.35$ ($Z_\mathrm{i}=0.008$);  (b) Padova \citep{marigo08} for $\mh=-0.375$ ($Z_\mathrm{i}=0.008$);  (c) MIST \citep{choi16} for $\mh=-0.37$ ($Z_\mathrm{i}=0.006$); (d): \parsec-\colibri\ isochrones (this work) for $\mh=-0.37$ ($Z_\mathrm{i}=0.006$); the heavy lines are for $\nintp=0$, the light ones for $\nintp=25$.}
\label{fig:comparison}
\end{figure*}

Owing to the numerical complexity and sizeable uncertainties in the evolution of TP-AGB stars, isochrones including this phase are still rare. Moreover, the few available sets of such isochrones are far from being similar -- as, instead, frequently occurs for evolutionary stages like the main sequence, RGB or CHeB. Figure~\ref{fig:comparison} presents a chronological sequence of such isochrones, zooming on the TP-AGB region of the HRD. The comparison is made for  metallicity values comparable to that of young populations in the LMC, namely $\mh=-0.35$~dex, which translates in slightly different values of $Z_\mathrm{i}$ in the different sets, depending on their definition of their solar value. A few differences are well evident in the figure:
\begin{itemize}
\item The \citet{cordier07} isochrones, which have extended the BaSTI database with the complete TP-AGB evolution, actually miss a lot of its details. The most evident difference the lack of a transition to the C-star phase in these isochrones. Moreover the slope of their TP-AGB sequences in the HRD is clearly different from the pre-AGB evolution. This is certainly due to the very basic synthetic TP-AGB modeling -- entirely based on fitting relations rather on envelope integrations -- they have adopted. Also the old Padova isochrones from \citet{girardi00} present a similar level of simplicity in their TP-AGB sections.
\item As already mentioned, the work by \citet{marigo08} represented a major effort to include detailed TP-AGB sections in isochrones. At first sight, their isochrones appear similar to present ones with $\nintp=0$. However, there are many different  details, like for instance: the nearly-absence of ``glitches'' in the present tables, due to the larger grids of tracks and improved interpolation schemes; the shorter extension in \Teff\ due to the more accurate low-temperature opacities; the different extensions to high luminosities in the younger isochrones, due to a more detailed treatment of HBB and mass-loss. Moreover, there are subtle differences in the distribution of stars along these isochrones, that do not appear in this particular figure; they have been discussed by \citet{marigo13,rosenfield14,rosenfield16}, and will not be repeated here. We also remark that similar TP-AGB sections are also found in the \citet{bertelli08,bertelli09} sets of isochrones.
\item The recent MIST isochrones \citep{choi16} include the TP-AGB phase computed with a full evolutionary code, together with its $L$-$\Teff$ variations due to TPCs. Their isochrone sequences appear more irregular than ours, owing to their more direct interpolation between TP-AGB tracks \citep[see][]{dotter16}. This aspect, however, should be seen as a less important -- and mostly aesthetic -- difference. The more important differences, instead, are in their C-star sections, which clearly start at a later point along the TP-AGB than ours, and for a decisively shorter age range. These features are very likely another manifestation of the old ``C-star mystery'' \citep[see][]{iben81,lattanzio04,herwig05} found in stellar evolution models which apply standard methods to model the extent of third dredge up events. Moreover, the C-rich sequences in MIST are in general hotter than in our isochrones, as a result of their use of molecular opacities derived for O-rich mixtures. 
\end{itemize}
Apart from the differences in the HRD, there will be other significant ones in diagrams relating other variables, such as the $\Mini$ which determines the occupation probability of every isochrone section.

\section{Concluding remarks}
\label{conclu}

As thoroughly discussed in \citet{marigo13} and \citet{rosenfield16}, the \colibri\ code models third dredge-up and mass-loss via a few simplified prescriptions. They contain parameters which, although reasonably constrained, have to be fine-tuned so as to reproduce the properties of TP-AGB star samples in nearby star clusters and galaxies, over the-widest-the-possible range of ages and metallicities. Moreover, all the grids of theoretical models involved in the construction of these isochrones are being expanded. This implies that present models will be frequently revised, in the context of the ERC-funded project STARKEY to calibrate the TP-AGB phase. Therefore, the present paper presents just a first snapshot of the isochrones involved in this process.
 
Isochrones presented in this paper, and in subsequent releases, can be retrieved from both the traditional CMD site \url{http://stev.oapd.inaf.it/cmd}, and from the new web server \url{http://starkey.astro.unipd.it/cgi-bin/cmd}. The servers provide isochrones computed on-the-fly, within seconds, for any sequence of ages and metallicities, and for a large set of pre-defined photometric systems. Additional computations -- like luminosity functions and integrated magnitudes of stellar populations in several passbands -- are also possible. Additional filters can be added upon request.

We note that massive stars/young isochrones are provided by the web interface as well, even if they do not regard the TP-AGB phase. Such massive star models are described in \citet{tang14} and \citet{chen15}, to whom we refer for all details.
 
The same isochrones are also being implemented as new alternatives in the \trilegal\  \citep{girardi05} code for simulating the photometry of resolved stellar populations, and in the \param\ code \citep{dasilva06, rodrigues14} for the Bayesian estimation of the properties of observed stars.
 
\acknowledgments
We acknowledge the support from the  ERC Consolidator Grant funding scheme ({\em project STARKEY}, G.A. n. 615604). LG and TSR acknowledge partial support from PRIN INAF 2014 -- CRA 1.05.01.94.05.

\bibliography{ms} 


\end{document}